\documentclass[11pt]{article}

\usepackage[T1]{fontenc}
\usepackage[utf8]{inputenc}
\usepackage{lmodern}
\usepackage{microtype}
\usepackage{mathtools,amssymb}
\usepackage[margin=1in]{geometry}
\usepackage{tikz}
\usepackage[font=small,labelfont=bf]{caption}
\usepackage{xurl}
\usepackage[hidelinks]{hyperref}

\usetikzlibrary{arrows.meta}

\numberwithin{equation}{section}
\allowdisplaybreaks[2]
\title{Expansion by Regions Derivation from the Mellin Transform}
\author{Andres Põldaru}
\date{}

\begin{document}

\maketitle

\begin{abstract}
We derive the method of expansion by regions from the Mellin transform for integrals involving polynomials with positive coefficients. This method is widely used to obtain asymptotic expansions of Feynman integrals.
\end{abstract}

\section{Introduction}

Let us consider an integral over \(n\) variables
\(\boldsymbol{x}=(x_1,\ldots,x_n)\), depending on a parameter \(t\),
of the form
\begin{equation}
	I(t;\boldsymbol{s}_x)
	=
	\int_{\mathbb{R}_+^n}
	\frac{d\boldsymbol{x}}{\boldsymbol{x}}\,
	\boldsymbol{x}^{\boldsymbol{s}_x}
	\boldsymbol{P}(\boldsymbol{x},t)^{\boldsymbol{\mu}}.
	\label{eq:introduction-integral}
\end{equation}
Here
\begin{equation}
	\boldsymbol{s}_x=(s_1,\ldots,s_n),
	\qquad
	\boldsymbol{x}^{\boldsymbol{s}_x}
	:=
	\prod_{i=1}^n x_i^{s_i},
	\qquad
	\frac{d\boldsymbol{x}}{\boldsymbol{x}}
	:=
	\prod_{i=1}^n\frac{dx_i}{x_i},
	\qquad
	\mathbb{R}_+^n=(0,\infty)^n,
	\qquad
	\boldsymbol{P}^{\boldsymbol{\mu}}
	:=
	\prod_{\ell=1}^M P_\ell^{\mu_\ell}.
\end{equation}
The components of
\(\boldsymbol{P}=(P_1,\ldots,P_M)\) are polynomials with positive
coefficients, while the components of
\(\boldsymbol{\mu}=(\mu_1,\ldots,\mu_M)\) are real exponents. Integrals
of such type can appear for example in Feynman integrals.

Generalizing the geometric method of expansion by regions
\cite{PakSmirnov2011}, its asymptotic series in \(t\) can be
obtained from the sum polytope of
\[
\Delta_{P_\ell}=\operatorname{Newt}(P_\ell),
\quad \ell=1,\ldots,M,
\qquad
\Delta_{\boldsymbol{P}}
=
\sum_{\ell=1}^M\Delta_{P_\ell}
\subset\mathbb{R}^{n+1},
\]
where the \((n+1)\)-st coordinate is the exponent of \(t\). We
take all inward pointing facet normal vectors of \(\Delta_{\boldsymbol{P}}\) whose final
component is positive and normalize them to the form
\[
\boldsymbol{u}_i=(\boldsymbol{\gamma}_i,1).
\]
For each such region vector, we rescale the integration variables
\[
x_j\longmapsto t^{\gamma_{i,j}}x_j,
\qquad j=1,\ldots,n,
\]
and expand the rescaled integrand in powers of \(t\). The expansion
is integrated over the full region
\(\boldsymbol{x}\in\mathbb{R}_+^n\). The sum of the expanded integrals over all region
vectors gives the asymptotic series of
\eqref{eq:introduction-integral} in \(t\). Alternatively, one
can introduce \(\lambda\), make the replacements
\(x_j\mapsto\lambda^{\gamma_{i,j}}x_j\) and \(t\mapsto\lambda t\), expand in
\(\lambda\), and set \(\lambda=1\) after the expansion to obtain the same
result, up to a rescaling of the integration variables.

For example, consider the integral
\begin{equation}
	I(t;s)
	=
	\int_0^\infty dx\,
	\frac{x^{s-1}}{1+x+t(1+x^3)}.
	\label{eq:introduction-region-convergence-example}
\end{equation}
For \(t>0\), the integral converges when
\[
	0<\operatorname{Re}s<3.
\]
The two relevant inward normals of the Newton polytope are
\((0,1)\) and \((-\tfrac12,1)\), giving the region replacements
\(x\mapsto x\) and \(x\mapsto t^{-1/2}x\), respectively.  The two
region expansions are
\begin{equation}
	I(t;s)
	\xrightarrow{x\mapsto x}
	\sum_{k=0}^{\infty}(-t)^k
	\int_0^\infty dx\,
	x^{s-1}\frac{(1+x^3)^k}{(1+x)^{k+1}},
	\label{eq:example-region-01}
\end{equation}
and
\begin{equation}
	\begin{split}
	I(t;s)
	&\xrightarrow{x\mapsto t^{-1/2}x}
	t^{(1-s)/2}
	\int_0^\infty dx\,
	\frac{x^{s-1}}{x(1+x^2)+t^{1/2}(1+t)}
	\\
	&\to
	t^{(1-s)/2}
	\sum_{k=0}^{\infty}
	(-1)^k t^{k/2}(1+t)^k
	\int_0^\infty dx\,
	\frac{x^{s-k-2}}{(1+x^2)^{k+1}}.
	\end{split}
	\label{eq:example-region-minus-half-1}
\end{equation}
In the first region, the \(k\)-th coefficient integral converges only
when
\[
	0<\operatorname{Re}s<1-2k.
\]
Thus its leading term, \(k=0\), converges for
\(0<\operatorname{Re}s<1\), whereas every coefficient integral with
\(k\geq1\) does not converge for any value of \(s\).  In practice, what is done is that for each individual integral, the integration space is divided into sectors and in each sector it is assumed that \(s\) is in such a range that this particular sector integral converges. Then the result is meromorphically continued to the value of \(s\) that we wanted in the original integral \eqref{eq:introduction-region-convergence-example} and using this method we would get the correct result for the series expansion in \(t\). However, it might not be clear why can we take undefined infinite integrals and convert them to some finite values, how do we know this is valid. In this article we will derive this method step by step and we will see why are we allowed to do it.

In the second region, the
\(k\)-th coefficient integral converges when
\[
	1+k<\operatorname{Re}s<3+3k.
\]
This means the higher order terms need increasingly larger values of \(s\) and there is no single value that makes them all converge.

\subsection{General Series Expansion}
To generalize the method of generating the series, keep the inward-pointing vector
\(\boldsymbol{u}_i=(\boldsymbol{\gamma}_i,1)\), and let
\(m_i\) be the least common multiple of the denominators of the components of
\(\boldsymbol{\gamma}_i\). After inserting \(\lambda\), all scaled
polynomials contain only integer powers of
\(\lambda^{1/m_i}\). For each \(\ell=1,\ldots,M\), let
\(n_{i,\ell}\) be the smallest power of \(\lambda^{1/m_i}\) in the
scaled polynomial \(P_\ell\), and set
\(\boldsymbol{n}_i=(n_{i,1},\ldots,n_{i,M})\). We factor these powers
out as
\begin{equation}
	P_\ell\!\left(
	\boldsymbol{x}\lambda^{\boldsymbol{\gamma}_i},
	\lambda t
	\right)
	=
	\lambda^{n_{i,\ell}/m_i}
	\widehat P_{i,\ell}(\lambda^{1/m_i}),
	\qquad \ell=1,\ldots,M.
\end{equation}
Thus \(n_{i,\ell}/m_i\) is the lowest power of \(\lambda\) in the
scaled \(P_\ell\), while \(\widehat P_{i,\ell}\) contains only
nonnegative integer powers of \(\lambda^{1/m_i}\).
We then do a formal Taylor expansion in
\(\lambda^{1/m_i}\), even though the series might not converge for all values of \(\boldsymbol{x}\):
\begin{align}
	&\lambda^{-\boldsymbol{\mu}\cdot\boldsymbol{n}_i/m_i}
	\boldsymbol{P}\!\left(
		\boldsymbol{x}\lambda^{\boldsymbol{\gamma}_i},
		\lambda t
	\right)^{\boldsymbol{\mu}}
	\notag\\
	&\qquad=
	\sum_{k=0}^{\infty}
	\frac{\lambda^{k/m_i}}{k!}
	\left(
	\left.
	\left(
		\frac{\partial}{\partial\lambda^{1/m_i}}
		\right)^k
		\left[
		\lambda^{-\boldsymbol{\mu}\cdot\boldsymbol{n}_i/m_i}
		\boldsymbol{P}\!\left(
			\boldsymbol{x}\lambda^{\boldsymbol{\gamma}_i},
			\lambda t
		\right)^{\boldsymbol{\mu}}
		\right]
		\right|_{\lambda=0}
		\right).
\end{align}
After setting \(\lambda=1\) in the prefactors, the full expanded integral associated
with the region vector \(\boldsymbol{u}_i\) is therefore
\begin{equation}
	I_i(t;\boldsymbol{s}_x)
	=
	\sum_{k=0}^{\infty}
	\frac{1}{k!}
	\int_{\mathbb{R}_+^n}
	\frac{d\boldsymbol{x}}{\boldsymbol{x}}\,
	\boldsymbol{x}^{\boldsymbol{s}_x}
	\left.
	\left(
		\frac{\partial}{\partial\lambda^{1/m_i}}
		\right)^k
		\left[
		\lambda^{-\boldsymbol{\mu}\cdot\boldsymbol{n}_i/m_i}
		\boldsymbol{P}\!\left(
			\boldsymbol{x}\lambda^{\boldsymbol{\gamma}_i},
			\lambda t
		\right)^{\boldsymbol{\mu}}
		\right]
		\right|_{\lambda=0}.
	\label{eq:introduction-single-region-full-integral}
\end{equation}

\subsection{Mellin Transform Introduction}
The method of expansion by regions can be derived from the Mellin
transform. Taking the Mellin transform in \(t\),
\[
\mathcal{M}[I](\boldsymbol{s}_x,s_{n+1})
=
\int_0^\infty dt\,
t^{s_{n+1}-1}I(t;\boldsymbol{s}_x),
\]
we can use methods similar to \cite{NilssonPassare2010,BerkeschForsgardPassare2014} to determine its poles
and residues. From the general theory of Mellin transforms the poles and residues of the meromorphic continuation of the Mellin transform directly give the series expansion in \(t\) (see,
for example,
\cite[Section~2, Theorem~4]{FlajoletGourdonDumas1995} and
\cite{FriotGreynatDeRafael2005}).

As a simple example, suppose that the Mellin transform of a
single-variable function \(f(x)\),
\[
\mathcal{M}(s)
=
\int_0^\infty dx\,x^{s-1}f(x),
\]
converges for \(-2<\operatorname{Re}s<1\) and its meromorphic continuation to \(-7<\operatorname{Re}s<1\) has only simple poles at \(s=-3\) and \(s=-5\), so in that
region it can be written as
\[
\mathcal{M}(s)
=
\frac{A(s)}{(s+3)(s+5)},
\]
where \(A(s)\) is holomorphic for \(-7<\operatorname{Re}s<1\).
The residues are
\[
\operatorname*{Res}_{s=-3}\mathcal{M}(s)
=\frac{A(-3)}{2},
\qquad
\operatorname*{Res}_{s=-5}\mathcal{M}(s)
=-\frac{A(-5)}{2}.
\]
From the general theory of Mellin transforms, the corresponding small-\(x\) expansion, including the remainder, is (assuming \(\mathcal{M}(\sigma+i\tau)=
O((1+|\tau|)^{-1-\varepsilon})\) with \(\varepsilon > 0\) as \(|\tau| \rightarrow \infty\))
\begin{align}
	f(x)
	&=
	\frac{A(-3)}{2}x^3
	-\frac{A(-5)}{2}x^5
	+
	\frac{1}{2\pi i}
	\int_{c-i\infty}^{c+i\infty}
	ds\,x^{-s}
	\frac{A(s)}{(s+3)(s+5)},
	\qquad c=-6,
	\notag\\
	&=
	\frac{A(-3)}{2}x^3
	-\frac{A(-5)}{2}x^5
	+O(x^6),
	\qquad x\to0.
	\label{eq:introduction-simple-mellin-example}
\end{align}

\section{A Mellin Transform Integral}

We take the Mellin transform of
\begin{equation}
	I(t;\boldsymbol{s}_x)
	=
	\int_{\mathbb{R}_+^n}
	\frac{d\boldsymbol{x}}{\boldsymbol{x}}\,
	\boldsymbol{x}^{\boldsymbol{s}_x}\,
	\boldsymbol{P}(\boldsymbol{x},t)^{\boldsymbol{\mu}}
	\label{eq:basic-integral}
\end{equation}
with respect
to \(t\), using \(s_{n+1}\) as the Mellin variable. We write
\(\boldsymbol{s}=(s_1,\ldots,s_n,s_{n+1})\), so that
\begin{equation}
	\mathcal{M}[I](\boldsymbol{s})
	=
	\int_{0}^{\infty} dt\, t^{s_{n+1}-1}
	\int_{\mathbb{R}_+^n}
	\frac{d\boldsymbol{x}}{\boldsymbol{x}}\,
	\boldsymbol{x}^{\boldsymbol{s}_x}\,
	\boldsymbol{P}(\boldsymbol{x},t)^{\boldsymbol{\mu}}
	=
	\int_{\mathbb{R}_+^{n+1}}
	\frac{d\boldsymbol{x}}{\boldsymbol{x}}\,
	\boldsymbol{x}^{\boldsymbol{s}}\,
	\boldsymbol{P}(\boldsymbol{x})^{\boldsymbol{\mu}}.
	\label{eq:mellin-transform}
\end{equation}
In the last equality we relabeled \(t=x_{n+1}\) and wrote
\(\boldsymbol{x}=(x_1,\ldots,x_n,x_{n+1})\).
This is an \((n+1)\)-dimensional Mellin transform, of which we need to find the poles and residues. We assume the integral converges for some value of \(\boldsymbol{s}\).

We now introduce logarithmic coordinates
\[
x_i=e^{y_i},
\qquad i=1,\ldots,n+1,
\]
and write \(\boldsymbol{y}=(y_1,\ldots,y_{n+1})\). Then the Mellin
transform becomes
\begin{equation}
	\mathcal{M}[I](\boldsymbol{s})
	=
	\int_{\mathbb{R}^{n+1}}
	d\boldsymbol{y}\,
	e^{\boldsymbol{s}\cdot\boldsymbol{y}}
	\boldsymbol{P}(e^{\boldsymbol{y}})^{\boldsymbol{\mu}}.
	\label{eq:mellin-log-coordinates}
\end{equation}
Here we used
\[
\frac{d\boldsymbol{x}}{\boldsymbol{x}}=d\boldsymbol{y},
\qquad
\boldsymbol{x}^{\boldsymbol{s}}
=e^{\boldsymbol{s}\cdot\boldsymbol{y}},
\]
and the integration domain \(\mathbb{R}_+^{n+1}\) became
\(\mathbb{R}^{n+1}\). For \(\ell=1,\ldots,M\), the polynomials have
the form
\[
P_\ell(e^{\boldsymbol{y}})
=
\sum_{a=1}^{N_\ell}c_{\ell,a}
e^{\boldsymbol{a}_{\ell,a}\cdot\boldsymbol{y}},
\]
where \(c_{\ell,a}>0\), while the exponent vectors
\(\boldsymbol{a}_{\ell,a}\) have integer components.

\subsection{Integration by Parts}

It is possible to reduce the scaling of the integrand in some direction by doing integration by parts in that direction. For a vector \(\boldsymbol{v}_i\) with integer components whose greatest common divisor is 1, define the largest exponents of
\(P_\ell\), for \(\ell=1,\ldots,M\), in that direction by
\[
p_{i,\ell}
=
\max_{1\leq a\leq N_\ell}
\boldsymbol{a}_{\ell,a}\cdot\boldsymbol{v}_i,
\qquad
\boldsymbol{p}_i
:=
(p_{i,1},\ldots,p_{i,M}).
\]
Choose linearly independent vectors
\(\boldsymbol{\eta}_{i1},\ldots,\boldsymbol{\eta}_{in}\) such that
\[
\boldsymbol{\eta}_{i1},\ldots,\boldsymbol{\eta}_{in},\boldsymbol{v}_i
\]
form a basis of \(\mathbb{R}^{n+1}\), and let
\[
W_i
:=
\operatorname{span}\left\{
\boldsymbol{\eta}_{i1},\ldots,\boldsymbol{\eta}_{in}
\right\}.
\]
We use the coordinates
\[
\boldsymbol{y}
=
\boldsymbol{w}_i+z_i\boldsymbol{v}_i,
\qquad
\boldsymbol{w}_i
=
\sum_{a=1}^n w_{ia}\boldsymbol{\eta}_{ia}
\in W_i.
\]
Define
\[
d^n\boldsymbol{w}_i
:=
\prod_{a=1}^n dw_{ia},
\qquad
J_i
:=
\left|
\det\left(
\boldsymbol{\eta}_{i1},\ldots,
\boldsymbol{\eta}_{in},\boldsymbol{v}_i
\right)
\right|.
\]
For this change of coordinates,
\begin{equation}
	d\boldsymbol{y}
	=
	J_i\,d^n\boldsymbol{w}_i\,d z_i,
	\qquad
	\partial_i
	:=
	\frac{\partial}{\partial z_i}
	=
	\boldsymbol{v}_i\cdot\nabla_{\boldsymbol{y}}.
	\label{eq:general-transverse-coordinate-jacobian}
\end{equation}
We define \(P_{i,\ell}\) by factoring out the largest exponential
power of each polynomial in the \(\boldsymbol{v}_i\)-direction:
\[
P_{i,\ell}(e^{\boldsymbol{y}})
:=
e^{-p_{i,\ell}z_i}P_\ell(e^{\boldsymbol{y}}),
\qquad \ell=1,\ldots,M.
\]
For convenience, set
\[
\boldsymbol{P}_i
:=
(P_{i,1},\ldots,P_{i,M}),
\qquad
\boldsymbol{1}
:=
(1,\ldots,1)\in\mathbb{R}^M,
\]
and use the same vector-power convention for
\(\boldsymbol{P}_i^{\boldsymbol{\mu}}\) as for
\(\boldsymbol{P}^{\boldsymbol{\mu}}\).
Consequently,
\[
\boldsymbol{P}^{\boldsymbol{\mu}}
=
e^{(\boldsymbol{\mu}\cdot\boldsymbol{p}_i)z_i}
\boldsymbol{P}_i^{\boldsymbol{\mu}}.
\]
The Mellin transform can therefore be written as
\[
\mathcal{M}[I]
=
J_i
	\int_{\mathbb{R}^n}d^n\boldsymbol{w}_i\,
	e^{\boldsymbol{s}\cdot\boldsymbol{w}_i}
	\int_{-\infty}^{\infty}d z_i\,
	e^{(\boldsymbol{s}\cdot\boldsymbol{v}_i
		+\boldsymbol{\mu}\cdot\boldsymbol{p}_i)z_i}
	\boldsymbol{P}_i^{\boldsymbol{\mu}}.
\]
Let
\[
\varphi_i(\boldsymbol{s})
:=
\boldsymbol{s}\cdot\boldsymbol{v}_i
+\boldsymbol{\mu}\cdot\boldsymbol{p}_i.
\]
Integrating by parts in \(z_i\) gives
\begin{align}
	\mathcal{M}[I]
	&=
	\frac{J_i}
	{\varphi_i(\boldsymbol{s})}
	\int_{\mathbb{R}^n}d^n\boldsymbol{w}_i\,
	e^{\boldsymbol{s}\cdot\boldsymbol{w}_i}
	\left[
	e^{\varphi_i(\boldsymbol{s})z_i}
	\boldsymbol{P}_i^{\boldsymbol{\mu}}
	\right]_{z_i=-\infty}^{z_i=\infty}
	\notag\\
	&\quad
	-\frac{J_i}
	{\varphi_i(\boldsymbol{s})}
	\int_{\mathbb{R}^n}d^n\boldsymbol{w}_i\,
	e^{\boldsymbol{s}\cdot\boldsymbol{w}_i}
	\int_{-\infty}^{\infty}d z_i\,
	e^{\varphi_i(\boldsymbol{s})z_i}
	\partial_i
	\left[
	\boldsymbol{P}_i^{\boldsymbol{\mu}}
	\right].
	\label{eq:log-coordinate-ibp}
\end{align}
Including the factor \(e^{\boldsymbol{s}\cdot\boldsymbol{w}_i}\), the
quantity in the boundary term is precisely the original integrand:
\[
e^{\boldsymbol{s}\cdot\boldsymbol{w}_i}
e^{\varphi_i(\boldsymbol{s})z_i}
\boldsymbol{P}_i^{\boldsymbol{\mu}}
=
e^{\boldsymbol{s}\cdot
	(\boldsymbol{w}_i+z_i\boldsymbol{v}_i)}
\boldsymbol{P}^{\boldsymbol{\mu}}.
\]
For the original integral to converge, this integrand is zero at the
boundaries \(z_i=\pm\infty\), so the boundary term vanishes.
Therefore we have
\begin{equation}
	\mathcal{M}[I]
	=
	-\frac{J_i}
	{\varphi_i(\boldsymbol{s})}
	\int_{\mathbb{R}^n}d^n\boldsymbol{w}_i\,
	e^{\boldsymbol{s}\cdot\boldsymbol{w}_i}
	\int_{-\infty}^{\infty}d z_i\,
	e^{\varphi_i(\boldsymbol{s})z_i}
	\partial_i
	\left[
	\boldsymbol{P}_i^{\boldsymbol{\mu}}
	\right].
	\label{eq:log-coordinate-ibp-operator-form}
\end{equation}
The derivative in the remaining integral is
\begin{equation}
	\partial_i
	\left(\boldsymbol{P}_i^{\boldsymbol{\mu}}\right)
	=
	\boldsymbol{P}_i^{\boldsymbol{\mu}}
	\sum_{\ell=1}^M
	\mu_\ell
	\frac{\partial_iP_{i,\ell}}{P_{i,\ell}}.
	\label{eq:log-coordinate-ratio-derivative}
\end{equation}

By construction, the highest \(e^{z_i}\)-power in every
\(P_{i,\ell}\) is zero. The derivative \(\partial_i\) removes every
term with power zero. Since all the \(e^{z_i}\)-powers are integers,
the highest power in \(\partial_iP_{i,\ell}\) is therefore at most
\(-1\). We can consequently write
\begin{equation}
	\partial_i
	\left(\boldsymbol{P}_i^{\boldsymbol{\mu}}\right)
	=
	e^{-z_i}
	\boldsymbol{P}_i^{\boldsymbol{\mu}}
	\frac{\widetilde{T}_{(i,1)}}
	{\boldsymbol{P}_i^{\boldsymbol{1}}},
	\label{eq:log-coordinate-ratio-derivative-factorized}
\end{equation}
where
\begin{equation}
	\widetilde{T}_{(i,1)}
	:=
	\sum_{\ell=1}^M
	\mu_\ell
	\bigl(e^{z_i}\partial_iP_{i,\ell}\bigr)
	\prod_{\substack{h=1\\h\neq\ell}}^M P_{i,h}.
	\label{eq:first-ibp-polynomial}
\end{equation}
Its highest \(e^{z_i}\)-power is either zero, or lower if the
next lower power of every \(P_\ell\) with \(\mu_\ell\neq0\) is at
least two below its highest power.

Integrating again by parts, the new boundary term is again proportional to the previous integrand and has to be zero at infinity for the previous integral to converge. After \(N_i\) integrations by parts in the
\(\boldsymbol{v}_i\)-direction, the integral is
\begin{align}
	\mathcal{M}[I]
	&=
	\frac{(-1)^{N_i}J_i}
	{\displaystyle
		\prod_{k=0}^{N_i-1}
		\bigl(\varphi_i(\boldsymbol{s})-k\bigr)}
	\int_{\mathbb{R}^n}d^n\boldsymbol{w}_i\,
	e^{\boldsymbol{s}\cdot\boldsymbol{w}_i}
	\int_{-\infty}^{\infty}d z_i\,
	e^{(\varphi_i(\boldsymbol{s})-N_i)z_i}
	\bigl(e^{z_i}\partial_i\bigr)^{N_i}
	\left[
	\boldsymbol{P}_i^{\boldsymbol{\mu}}
	\right]
	\label{eq:fixed-direction-Ni-integrations-operator}\\
	&=
	\frac{(-1)^{N_i}J_i}
	{\displaystyle
		\prod_{k=0}^{N_i-1}
		\bigl(\varphi_i(\boldsymbol{s})-k\bigr)}
	\int_{\mathbb{R}^n}d^n\boldsymbol{w}_i\,
	e^{\boldsymbol{s}\cdot\boldsymbol{w}_i}
	\int_{-\infty}^{\infty}d z_i\,
	e^{(\varphi_i(\boldsymbol{s})-N_i)z_i}
	\boldsymbol{P}_i^{\boldsymbol{\mu}}
	\frac{\widetilde{T}_{(i,N_i)}}
	{\boldsymbol{P}_i^{N_i\boldsymbol{1}}}.
	\label{eq:fixed-direction-Ni-integrations}
\end{align}
where
\(\widetilde{T}_{(i,N_i)}\) is defined recursively by
\begin{equation}
	e^{z_i}\partial_i
	\left[
	\boldsymbol{P}_i^{\boldsymbol{\mu}}
	\frac{\widetilde{T}_{(i,N_i)}}
	{\boldsymbol{P}_i^{N_i\boldsymbol{1}}}
	\right]
	=
	\boldsymbol{P}_i^{\boldsymbol{\mu}}
	\frac{\widetilde{T}_{(i,N_i+1)}}
	{\boldsymbol{P}_i^{(N_i+1)\boldsymbol{1}}}.
	\label{eq:fixed-direction-tilde-T-operator-action}
\end{equation}
This gives
\begin{align}
	\widetilde{T}_{(i,N_i+1)}
	:={}&
	\boldsymbol{P}_i^{\boldsymbol{1}}
	\bigl(e^{z_i}\partial_i\widetilde{T}_{(i,N_i)}\bigr)
	\notag\\
	&+
	\sum_{\ell=1}^M
	(\mu_\ell-N_i)
	\bigl(e^{z_i}\partial_iP_{i,\ell}\bigr)
	\prod_{\substack{h=1\\h\neq\ell}}^M P_{i,h}\,
	\widetilde{T}_{(i,N_i)}.
	\label{eq:fixed-direction-tilde-T-recursion}
\end{align}
From this recursion we see that
\(\widetilde{T}_{(i,N_i)}\) is a sum of terms of the form
\begin{equation}
	C^{(i,N_i)}_{\boldsymbol{\alpha}_1,\ldots,
		\boldsymbol{\alpha}_M}(\boldsymbol{\mu})
	\prod_{\ell=1}^M
	\prod_{a=1}^{N_i}
	\left[
	\bigl(e^{z_i}\partial_i\bigr)^{\alpha_{\ell,a}}
	P_{i,\ell}
	\right],
	\label{eq:fixed-direction-tilde-T-general-term}
\end{equation}
where \(\alpha_{\ell,a}\in\mathbb{Z}_{\geq0}\) and
\begin{equation}
	\sum_{\ell=1}^M
	\sum_{a=1}^{N_i}\alpha_{\ell,a}
	=N_i.
	\label{eq:fixed-direction-tilde-T-operator-count}
\end{equation}
Thus each term contains exactly \(N_i\) factors built from each
\(P_{i,\ell}\), in total \(MN_i\) factors, while the total number of
\(e^{z_i}\partial_i\) applications is \(N_i\). The highest \(e^{z_i}\)-power of each term and therefore of \(\widetilde{T}_{(i,N_i)}\) is zero or lower. The coefficients
\(C^{(i,N_i)}_{\boldsymbol{\alpha}_1,\ldots,
\boldsymbol{\alpha}_M}(\boldsymbol{\mu})\) are determined by the
recursion above.

The same terms, with different coefficients, are produced by the
following polynomial
\begin{equation}
	\widetilde{T}_{(i,N_i)}^*
	:=
	\bigl(e^{z_i}\partial_i\bigr)^{N_i}
	\left[
	\boldsymbol{P}_i^{N_i\boldsymbol{1}}
	\right].
	\label{eq:fixed-direction-tilde-T-scaling-replacement}
\end{equation}
Indeed, by the Leibniz rule, the \(N_i\) derivatives are distributed
among the \(N_i\) factors of every \(P_{i,\ell}\) in all ways
satisfying
Eq.~\eqref{eq:fixed-direction-tilde-T-operator-count}.
We write
\begin{equation}
	\widetilde{T}_{(i,N_i)}
	\cong
	\widetilde{T}_{(i,N_i)}^*
\end{equation}
to mean that the two polynomials have the same exponential
scaling before coefficient cancellations. Some coefficients in
\(\widetilde{T}_{(i,N_i)}\) may vanish, but this can only remove powers
and hence lower its exponential scaling; it cannot increase it. Thus
\(\widetilde{T}_{(i,N_i)}^*\) provides an upper bound on the exponential
scaling of \(\widetilde{T}_{(i,N_i)}\).

If we now write the integral using the original \(P_\ell\),
without factoring out their largest powers of \(e^{z_i}\), then
\begin{equation}
	\mathcal{M}[I]=
	\frac{(-1)^{N_i}J_i}
	{\displaystyle
		\prod_{k=0}^{N_i-1}
		\bigl(\varphi_i(\boldsymbol{s})-k\bigr)}
	\int_{\mathbb{R}^n}d^n\boldsymbol{w}_i\,
	e^{\boldsymbol{s}\cdot\boldsymbol{w}_i}
	\int_{-\infty}^{\infty}dz_i\,
	e^{(\boldsymbol{s}\cdot\boldsymbol{v}_i)z_i}
	\boldsymbol{P}\!\left(
		e^{\boldsymbol{w}_i+z_i\boldsymbol{v}_i}
	\right)^{\boldsymbol{\mu}}
	\frac{T_{(i,N_i)}}{
		\boldsymbol{P}\!\left(
			e^{\boldsymbol{w}_i+z_i\boldsymbol{v}_i}
		\right)^{N_i\boldsymbol{1}}},
	\label{eq:fixed-direction-factorized-and-original-integrands}
\end{equation}
where
\begin{equation}
	T_{(i,N_i)}
	:=
	e^{-N_i(1-\boldsymbol{1}\cdot\boldsymbol{p}_i)z_i}
	\widetilde{T}_{(i,N_i)}.
	\label{eq:fixed-direction-tilde-T-to-T}
\end{equation}
The corresponding
congruent form of \(T_{(i,N_i)}\) is
\begin{equation}
	T_{(i,N_i)}
	\cong
	e^{N_i \boldsymbol{1}\cdot\boldsymbol{p}_i z_i}e^{-N_iz_i}
	\bigl(e^{z_i}\partial_i\bigr)^{N_i}
	\left[
	e^{-N_i \boldsymbol{1}\cdot\boldsymbol{p}_i z_i}
	\boldsymbol{P}^{N_i\boldsymbol{1}}
	\right].
	\label{eq:fixed-direction-T-scaling-replacement}
\end{equation}

We can convert the \(z_i\) derivatives to derivatives with respect to an auxiliary variable \(\tau_i\).
Because the \(z_i\)-integration extends over the whole real line, we may make the translation
\begin{equation}
	z_i\longmapsto z_i+\tau_i
\end{equation}
For every differentiable function \(f\),
\begin{equation}
	\frac{\partial}{\partial z_i}f(z_i+\tau_i)
	=
	\frac{\partial}{\partial\tau_i}f(z_i+\tau_i).
\end{equation}
Thus, after the translation, derivatives with respect to \(z_i\) may
be replaced by derivatives with respect to \(\tau_i\); in particular,
\begin{equation}
	e^{z_i}\partial_i
	\longmapsto
	e^{z_i+\tau_i}\frac{\partial}{\partial\tau_i}.
\end{equation}
Then all factors involving \(z_i\) can be
pulled in front and the integral becomes
\begin{align}
	\mathcal{M}[I]
	={}&
	\frac{(-1)^{N_i}}
	{\displaystyle
		\prod_{k=0}^{N_i-1}
		\bigl(\varphi_i(\boldsymbol{s})-k\bigr)}
	\int_{\mathbb{R}^{n+1}}d\boldsymbol{y}\,
	e^{\boldsymbol{s}\cdot\boldsymbol{y}}
	e^{(\varphi_i(\boldsymbol{s})-N_i)\tau_i}
	\notag\\
	&\quad\times
	\left(
	e^{\tau_i}\frac{\partial}{\partial\tau_i}
	\right)^{N_i}
	\left[
	e^{-(\boldsymbol{\mu}\cdot\boldsymbol{p}_i)\tau_i}
	\boldsymbol{P}\!\left(
		e^{\boldsymbol{y}+\tau_i\boldsymbol{v}_i}
	\right)^{\boldsymbol{\mu}}
	\right].
	\label{eq:fixed-direction-shifted-derivative-form}
\end{align}
After applying the derivatives, \(\tau_i\) may be assigned any fixed
value inside the integral; for example, we may set \(\tau_i=0\).

If we now repeat the integration by parts \(N_j\) times in another direction
\(\boldsymbol{v}_j\), with \(j\neq i\), and set \(N=N_i+N_j\), then
\begin{align}
	\mathcal{M}[I]
	={}&
	\frac{(-1)^N}
	{\displaystyle
		\prod_{k=0}^{N_i-1}
		\bigl(\varphi_i(\boldsymbol{s})-k\bigr)
		\prod_{\ell=0}^{N_j-1}
		\bigl(\varphi_j(\boldsymbol{s})-\ell\bigr)}
	\int_{\mathbb{R}^{n+1}}d\boldsymbol{y}\,
	e^{\boldsymbol{s}\cdot\boldsymbol{y}}
	e^{(\varphi_i(\boldsymbol{s})-N_i)\tau_i
		+(\varphi_j(\boldsymbol{s})-N_j)\tau_j}
	\notag\\
	&\quad\times
	\left(
	e^{\tau_j}\frac{\partial}{\partial\tau_j}
	\right)^{N_j}
	\left(
	e^{\tau_i}\frac{\partial}{\partial\tau_i}
	\right)^{N_i}
	\left[
	e^{-(\boldsymbol{\mu}\cdot\boldsymbol{p}_i)\tau_i
		-(\boldsymbol{\mu}\cdot\boldsymbol{p}_j)\tau_j}
	\boldsymbol{P}\!\left(
		e^{\boldsymbol{y}
			+\tau_i\boldsymbol{v}_i
			+\tau_j\boldsymbol{v}_j}
	\right)^{\boldsymbol{\mu}}
	\right].
	\label{eq:two-direction-shifted-operator}
\end{align}
And so on if we integrate by parts in a third direction. Note that the operators using different \(\tau_i\) commute. In general, let \(N_i\) be the number of integrations by parts in the direction
\(\boldsymbol{v}_i\) and \(L\) the total number of different directions we integrated in. Let
\begin{equation}
	N
	:=
	\sum_{i=1}^{L}N_i,
\end{equation}
and write
\(\boldsymbol{N}=(N_1,\ldots,N_L)\). After integrating by parts in all
facet-normal directions, we obtain
\begin{equation}
	\mathcal{M}[I]
	=
	\frac{(-1)^N}
	{\displaystyle
		\prod_{i=1}^{L}
		\prod_{k=0}^{N_i-1}
		\bigl(\varphi_i(\boldsymbol{s})-k\bigr)}
	\int_{\mathbb{R}^{n+1}}d\boldsymbol{y}\,
	e^{\boldsymbol{s}\cdot\boldsymbol{y}}
	\boldsymbol{P}^{\boldsymbol{\mu}}
	\frac{T_{\boldsymbol{N}}}
	{\boldsymbol{P}^{N\boldsymbol{1}}}.
	\label{eq:all-direction-integrations}
\end{equation}
Here products corresponding to \(N_i=0\) are understood as empty
products. Generalizing
Eq.~\eqref{eq:fixed-direction-T-scaling-replacement}, the resulting
numerator polynomial has the congruent form
\begin{align}
	T_{\boldsymbol{N}}(e^{\boldsymbol{y}})
	\cong
	T_{\boldsymbol{N}}^*(e^{\boldsymbol{y}})
	:={}&
	\exp\left(
	\sum_{i=1}^{L}
	\bigl(
	N(\boldsymbol{1}\cdot\boldsymbol{p}_i)-N_i
	\bigr)\tau_i
	\right)
	\left[
	\prod_{i=1}^{L}
	\left(
	e^{\tau_i}\frac{\partial}{\partial\tau_i}
	\right)^{N_i}
	\right]
	\notag\\
	&\quad\times
	\left[
	\exp\left(
	-N\sum_{i=1}^{L}
	(\boldsymbol{1}\cdot\boldsymbol{p}_i)\tau_i
	\right)
	\boldsymbol{P}\!\left(
		e^{\boldsymbol{y}+\sum_{i=1}^{L}
			\tau_i\boldsymbol{v}_i}
	\right)^{N\boldsymbol{1}}
	\right].
	\label{eq:all-direction-numerator-tau-derivatives}
\end{align}
The Mellin transform itself can equivalently be written directly in
terms of the auxiliary-variable derivatives as
\begin{align}
	\mathcal{M}[I]
	={}&
	\frac{(-1)^N}
	{\displaystyle
		\prod_{i=1}^{L}
		\prod_{k=0}^{N_i-1}
		\bigl(\varphi_i(\boldsymbol{s})-k\bigr)}
	\int_{\mathbb{R}^{n+1}}d\boldsymbol{y}\,
	e^{\boldsymbol{s}\cdot\boldsymbol{y}}
	\exp\left(
	\sum_{i=1}^{L}
	\bigl(\varphi_i(\boldsymbol{s})-N_i\bigr)\tau_i
	\right)
	\notag\\
	&\quad\times
	\left[
	\prod_{i=1}^{L}
	\left(
	e^{\tau_i}\frac{\partial}{\partial\tau_i}
	\right)^{N_i}
	\right]
	\left[
	\exp\left(
	-\sum_{i=1}^{L}
	(\boldsymbol{\mu}\cdot\boldsymbol{p}_i)\tau_i
	\right)
	\boldsymbol{P}\!\left(
		e^{\boldsymbol{y}+\sum_{i=1}^{L}
			\tau_i\boldsymbol{v}_i}
	\right)^{\boldsymbol{\mu}}
	\right].
	\label{eq:all-direction-shifted-operator}
\end{align}

\subsection{Convergence}

For the types of integrals in Eq.~\eqref{eq:mellin-log-coordinates} to converge, the exponential scaling of the integrand in any direction has to be negative. The exponential scaling can be easily determined in an integration sector where the dominant vertex of each of the polynomials in \(\boldsymbol{P}\) stays the same. For this we look at cones generated by \(n+1\) facet normal vectors of the Newton polytope \(\Delta_{\boldsymbol{P}}\) (full-dimensional Newton polytope of the product of the polynomials). A combination of such cones together cover the whole integration space and in each cone, looking at one polynomial, the same monomial stays with the largest exponent. The cones are called maximal cones of a simplicial refinement of the normal fan.

One such cone has the
form
\begin{equation}
	C_\sigma
	=
	\left\{
	\boldsymbol{y}
	=
	\sum_{i\in I_\sigma}\xi_i\boldsymbol{v}_i:
	\xi_i\geq0
	\right\},
	\label{eq:normal-cone}
\end{equation}
where the \(n+1\) vectors \(\boldsymbol{v}_i\), \(i\in I_\sigma\), are
outwards pointing linearly independent facet normal vectors of \(\Delta_{\boldsymbol{P}}\). Let the total number of facet normal vectors be \(L\).

Every exponent vector \(\boldsymbol{a}\) in
the polytope \(N\Delta_{\boldsymbol{P}}\) satisfies
\begin{equation}
	\boldsymbol{a}\cdot\boldsymbol{v}_i
	\leq
	N(\boldsymbol{1}\cdot\boldsymbol{p}_i),
	\qquad i=1,\ldots,L.
	\label{eq:P-newton-polytope-cone-inequalities}
\end{equation}
Here the components of \(\boldsymbol{p}_i\) are the extremal powers
of the polynomials \(P_\ell\) in the
\(\boldsymbol{v}_i\)-direction:
\begin{equation}
	p_{i,\ell}
	=
	\max_{\boldsymbol{a}\in\Delta_{P_\ell}}
	\boldsymbol{a}\cdot\boldsymbol{v}_i,
	\qquad
	i=1,\ldots,L,\quad \ell=1,\ldots,M.
\end{equation}
On the interior of
\(C_\sigma\), a single vertex
\(\boldsymbol{a}_{\sigma,\ell}\) of each
\(\Delta_{P_\ell}\) has the largest power. Their sum is the
corresponding vertex of \(\Delta_{\boldsymbol{P}}\), and
\[
\boldsymbol{a}_{\sigma,\ell}\cdot\boldsymbol{v}_i
=p_{i,\ell},
\qquad
i\in I_\sigma,\quad \ell=1,\ldots,M.
\]
We can therefore factor the largest powers out as
\begin{equation}
	P_\ell(e^{\boldsymbol{y}})
	=
	e^{\boldsymbol{a}_{\sigma,\ell}\cdot\boldsymbol{y}}
	P_{\sigma,\ell}(\boldsymbol{y}),
	\qquad \ell=1,\ldots,M.
	\label{eq:dominant-vertex-factorization}
\end{equation}
Because all coefficients of the \(P_\ell\) are positive, the
factorized functions are bounded above and away from zero on the cone:
\[
0<c_{\sigma,\ell}
\leq P_{\sigma,\ell}
\leq C_{\sigma,\ell},
\qquad \ell=1,\ldots,M.
\]
Write \(I_\sigma=\{i_1,\ldots,i_{n+1}\}\) and set
\[
J_\sigma
:=
\left|
\det\left(
\boldsymbol{v}_{i_1},\ldots,\boldsymbol{v}_{i_{n+1}}
\right)
\right|,
\qquad
\boldsymbol{P}_\sigma
:=
(P_{\sigma,1},\ldots,P_{\sigma,M}).
\]
In the cone coordinates of Eq.~\eqref{eq:normal-cone}, the contribution
of \(C_\sigma\) to the Mellin integral is
\begin{align}
	\mathcal{M}_\sigma[I]
	={}&
	J_\sigma
	\int_{\mathbb{R}_+^{n+1}}
	\left(
	\prod_{i\in I_\sigma}d\xi_i
	\right)
	\exp\left(
	\sum_{i\in I_\sigma}
	\bigl(
		\boldsymbol{s}\cdot\boldsymbol{v}_i
		+\boldsymbol{\mu}\cdot\boldsymbol{p}_i
	\bigr)\xi_i
	\right)
	\notag\\
	&\quad\times
	\boldsymbol{P}_\sigma
	\left(
	\sum_{i\in I_\sigma}\xi_i\boldsymbol{v}_i
	\right)^{\boldsymbol{\mu}}.
	\label{eq:original-mellin-integral-on-cone}
\end{align}
The last factor is bounded above and away from zero on
\(C_\sigma\). Hence this cone integral converges if
\begin{equation}
	\operatorname{Re}\!\left(
	\boldsymbol{s}\cdot\boldsymbol{v}_i
	+\boldsymbol{\mu}\cdot\boldsymbol{p}_i
	\right)
	=
	\operatorname{Re}\varphi_i(\boldsymbol{s})
	<0,
	\qquad i\in I_\sigma.
	\label{eq:original-cone-convergence}
\end{equation}

Therefore, considering all such cones, the convergence region of the Mellin integral
\eqref{eq:mellin-log-coordinates} is determined by the inequalities
\begin{equation}
	\operatorname{Re}\!\left(
	\boldsymbol{s}\cdot\boldsymbol{v}_i
	+\boldsymbol{\mu}\cdot\boldsymbol{p}_i
	\right)
	=
	\operatorname{Re}\varphi_i(\boldsymbol{s})
	<0,
	\qquad i=1,\ldots,L.
\end{equation}

\subsubsection{Increasing the convergence region}

To increase the convergence region, we can do integration by parts. Note that integrating by parts in a direction that is not an outwards pointing
facet normal does not, in general, enlarge the convergence region,
because the inequalities associated with the facet normals remain in
force. For example, consider
\begin{equation}
	\int_{\mathbb{R}^2}dy_1\,dy_2\,
	\frac{e^{s_1y_1+s_2y_2}}
	{1+e^{y_1}+e^{y_2}}.
\end{equation}
The facet-normal directions \((0,-1)\), \((-1,0)\), and \((1,1)\)
give, respectively, the convergence conditions
\begin{equation}
	\operatorname{Re}s_2>0,
	\qquad
	\operatorname{Re}s_1>0,
	\qquad
	\operatorname{Re}(s_1+s_2)<1.
\end{equation}
The non-facet direction \((1,0)\) separately gives the redundant
condition \(\operatorname{Re}s_1<1\). One integration by parts in that
direction relaxes this directional condition to
\(\operatorname{Re}s_1<2\), but it does not enlarge the full convergence
region: the unchanged facet inequalities \(\operatorname{Re}s_2>0\)
and \(\operatorname{Re}(s_1+s_2)<1\) still imply
\(\operatorname{Re}s_1<1\). This illustrates why the
integrations by parts used to enlarge the convergence region are
performed along the outwards pointing facet-normal directions.

After integration by parts, the integrand has an extra term
\(\frac{T_{\boldsymbol{N}}}{\boldsymbol{P}^{N\boldsymbol{1}}}\).
To determine the
exponential scaling of \(T_{\boldsymbol{N}}\) in any direction, we may as an upper bound use the congruent form \(T_{\boldsymbol{N}}^*\)
derived in Eq.~\eqref{eq:all-direction-numerator-tau-derivatives}.

In that equation the operator
\[
\exp\left(
\bigl(N(\boldsymbol{1}\cdot\boldsymbol{p}_i)-N_i\bigr)\tau_i
\right)
\left(
e^{\tau_i}\frac{\partial}{\partial\tau_i}
\right)^{N_i}
\exp\left(
-N(\boldsymbol{1}\cdot\boldsymbol{p}_i)\tau_i
\right)
\]
for
each \(i=1,\ldots,L\) removes highest powers in the
\(\boldsymbol{v}_i\)-direction, from
\(N(\boldsymbol{1}\cdot\boldsymbol{p}_i)\) down to
\(N(\boldsymbol{1}\cdot\boldsymbol{p}_i)-N_i+1\). Therefore, after
\(N_i\) integrations by parts in the \(\boldsymbol{v}_i\)-direction \(T_{\boldsymbol{N}}\)
satisfies
\begin{equation}
	\max_{\boldsymbol{a}\in
		\operatorname{Newt}(T_{\boldsymbol{N}})}
	\boldsymbol{a}\cdot\boldsymbol{v}_i
	\leq
	N(\boldsymbol{1}\cdot\boldsymbol{p}_i)-N_i.
	\label{eq:T-star-highest-power-constraint}
\end{equation}

For one cone, we can obtain an upper bound on the largest power of \(T_{\boldsymbol{N}}\) in the direction
\(\boldsymbol{y}=\sum_{i\in I_\sigma}
\xi_i\boldsymbol{v}_i\):
\begin{align}
	\max_{\boldsymbol{a}\in
		\operatorname{Newt}(T_{\boldsymbol{N}})}\boldsymbol{a}\cdot\boldsymbol{y}
	&=
	\max_{\boldsymbol{a}\in
		\operatorname{Newt}(T_{\boldsymbol{N}})}
		\left(
		\sum_{i\in I_\sigma}
		\xi_i\boldsymbol{a}\cdot\boldsymbol{v}_i
		\right)
	\notag\\
	&\leq
	\sum_{i\in I_\sigma}
		\max_{\boldsymbol{a}\in
			\operatorname{Newt}(T_{\boldsymbol{N}})}
		\left(
		\xi_i\boldsymbol{a}\cdot\boldsymbol{v}_i
		\right)
	\notag\\
	&=
	\sum_{i\in I_\sigma}
	\xi_i
	\max_{\boldsymbol{a}\in
		\operatorname{Newt}(T_{\boldsymbol{N}})}
	\boldsymbol{a}\cdot\boldsymbol{v}_i
	\notag\\
	&\leq
	\sum_{i\in I_\sigma}
	\bigl(
	N(\boldsymbol{1}\cdot\boldsymbol{p}_i)-N_i
	\bigr)\xi_i.
	\label{eq:T-highest-power-on-cone}
\end{align}
It
follows that inside the cone \(C_\sigma\), there is some constant \(C_{T_{\boldsymbol{N}},\sigma}\), such that
\begin{equation}
	\left|T_{\boldsymbol{N}}\right|
	\leq
	C_{T_{\boldsymbol{N}},\sigma}
	\exp\left(
	\sum_{i\in I_\sigma}
	\bigl(
	N(\boldsymbol{1}\cdot\boldsymbol{p}_i)-N_i
	\bigr)\xi_i
	\right).
\end{equation}

Multiplying this estimate by the highest powers from the remaining factor
\(e^{\boldsymbol{s}\cdot\boldsymbol{y}}
\frac{\boldsymbol{P}^{\boldsymbol{\mu}}}
{\boldsymbol{P}^{N\boldsymbol{1}}}\), the absolute value of the integrand is bounded
by a constant times
\begin{align}
	&\exp\left(
	\sum_{i\in I_\sigma}
		\bigl(
			\operatorname{Re}\boldsymbol{s}\cdot\boldsymbol{v}_i
			+\boldsymbol{\mu}\cdot\boldsymbol{p}_i-N_i
		\bigr)\xi_i
		\right)
	\notag\\
	&\qquad=
	\exp\left(
	\sum_{i\in I_\sigma}
	\bigl(\operatorname{Re}\varphi_i(\boldsymbol{s})-N_i\bigr)
	\xi_i
	\right).
	\label{eq:cone-integrand-bound}
\end{align}
The Jacobian for \(\boldsymbol{y}=\sum_{i\in I_\sigma}\xi_i\boldsymbol{v}_i\) is
constant. Therefore the integral over \(C_\sigma\) converges absolutely
when
\begin{equation}
	\operatorname{Re}\varphi_i(\boldsymbol{s})<N_i,
	\qquad i\in I_\sigma.
	\label{eq:cone-convergence-after-ibp}
\end{equation}
Indeed, under these inequalities the total exponent is negative in
every generating direction of \(C_\sigma\). Therefore there is
a constant \(\delta_\sigma>0\) such that the absolute value of the
integrand is bounded by
\begin{equation}
	K_\sigma
	\exp\left(
	-\delta_\sigma\sum_{i\in I_\sigma}\xi_i
	\right).
\end{equation}
To see this explicitly, let
	\(\varepsilon_\sigma=
	\min_{i\in I_\sigma}
	\bigl(N_i-\operatorname{Re}\varphi_i(\boldsymbol{s})\bigr)>0\).
Then
\[
	\sum_{i\in I_\sigma}
	\bigl(N_i-\operatorname{Re}\varphi_i(\boldsymbol{s})\bigr)\xi_i
	\geq
	\varepsilon_\sigma\sum_{i\in I_\sigma}\xi_i.
\]
We may therefore take
\(\delta_\sigma=\varepsilon_\sigma\). Writing \(J_\sigma\) for the
constant Jacobian of the cone coordinates, we obtain
\begin{equation}
	J_\sigma K_\sigma
	\int_{\mathbb{R}_+^{n+1}}
	\left(\prod_{i\in I_\sigma}d\xi_i\right)
	e^{-\delta_\sigma\sum_{i\in I_\sigma}\xi_i}
	=
	\frac{J_\sigma K_\sigma}{\delta_\sigma^{n+1}}
	<\infty.
\end{equation}
The full integral converges when these inequalities hold on all such cones. Thus each integration by parts in direction
\(\boldsymbol{v}_i\) relaxes the inequalities in \(\boldsymbol{s}\) for the integrand to converge by at least one in the corresponding direction:
\begin{equation}
	\operatorname{Re}\varphi_i(\boldsymbol{s})<N_i,
	\qquad i=1,\ldots,L.
	\label{eq:full-convergence-after-ibp}
\end{equation}
By doing repeated integration by parts along all normal vectors of \(\Delta_{\boldsymbol{P}}\), we can make the integral's convergence region in \(\boldsymbol{s}\) arbitrarily large.

\subsubsection{Decay along the imaginary axis}
We can demonstrate that the growth condition~(23) in
\cite[Section~2, Theorem~4]{FlajoletGourdonDumas1995} holds, which is required to get the asymptotic series from the Mellin transform. The condition on 
\(\mathcal{M}[I](\boldsymbol{s}_x,s_{n+1})\) is that when \(s_{n+1}=\sigma+i\tau\) we need
\(\mathcal{M}[I](\boldsymbol{s}_x,\sigma+i\tau)=O((1+|\tau|)^{-1-\varepsilon})\), with
\(\varepsilon>0\) as \(|\tau|\to\infty\).

We can perform enough integrations by parts
that the remaining integral converges in an arbitrarily large region, including the range that is used in proving \cite[Section~2, Theorem~4]{FlajoletGourdonDumas1995}.

In particular, we do at least two integrations in a facet-normal direction
\(\boldsymbol{v}_i=(\boldsymbol{v}_{i,x},q_i)\), for which \(q_i\neq0\). These two integrations produce the prefactor
\begin{equation}
	\frac{1}{
		\varphi_i(\boldsymbol{s})
		\bigl(\varphi_i(\boldsymbol{s})-1\bigr)}.
\end{equation}
Writing \(s_{n+1}=\sigma+i\tau\), each of the two corresponding denominator factors
satisfies
\begin{equation}
	\left|\varphi_i(\boldsymbol{s})-k\right|
	=
	|q_i\tau|+O(1),
	\qquad |\tau|\to\infty.
\end{equation}
The remaining integral without the prefactor has the form
\begin{equation}
	\mathcal{I}_{\boldsymbol{N}}(\boldsymbol{s})
	:=
	\int_{\mathbb{R}^{n+1}}d\boldsymbol{y}\,
	e^{\boldsymbol{s}\cdot\boldsymbol{y}}
	\boldsymbol{P}(e^{\boldsymbol{y}})^{\boldsymbol{\mu}}
	\frac{T_{\boldsymbol{N}}(e^{\boldsymbol{y}})}
	{\boldsymbol{P}(e^{\boldsymbol{y}})^{N\boldsymbol{1}}}.
\end{equation}
It is uniformly bounded because
\begin{equation}
	\left|\mathcal{I}_{\boldsymbol{N}}(\boldsymbol{s})\right|
	\leq
	\int_{\mathbb{R}^{n+1}}d\boldsymbol{y}\,
	e^{\operatorname{Re}\boldsymbol{s}\cdot\boldsymbol{y}}
	\left|
	\boldsymbol{P}(e^{\boldsymbol{y}})^{\boldsymbol{\mu}}
	\frac{T_{\boldsymbol{N}}(e^{\boldsymbol{y}})}
	{\boldsymbol{P}(e^{\boldsymbol{y}})^{N\boldsymbol{1}}}
	\right|,
\end{equation}
and the convergent integral on the right is independent of \(\tau\).
Consequently,
\begin{equation}
	\mathcal{M}[I](\boldsymbol{s}_x,\sigma+i\tau)
	=
	O\!\left(|\tau|^{-2}\right),
	\qquad |\tau|\to\infty,
\end{equation}
which is more than is required. Doing more integrations by parts would make the scaling exponent in \(|\tau|\) even smaller.

\subsection{Poles and Residues}

Looking at the generalized expression after integration by parts in
Eq.~\eqref{eq:all-direction-shifted-operator}, the possible pole
hyperplanes produced by the denominator factors are
\begin{equation}
	\varphi_i(\boldsymbol{s})-k
	=\boldsymbol{s}\cdot\boldsymbol{v}_i
	+\boldsymbol{\mu}\cdot\boldsymbol{p}_i-k=0,
	\qquad
	i=1,\ldots,L,
	\qquad
	k\in\mathbb{Z}_{\geq0}.
	\label{eq:all-pole-hyperplanes}
\end{equation}
Write
\[
\boldsymbol{v}_i
=
(\boldsymbol{v}_{i,x},q_i),
\qquad
q_i=(\boldsymbol{v}_i)_{n+1}.
\]
If \(q_i\neq0\), the corresponding poles in \(s_{n+1}\) are
\begin{equation}
	s_{p,i,k}(\boldsymbol{s}_x)
	=
	\frac{
		k-\boldsymbol{s}_x\cdot\boldsymbol{v}_{i,x}
		-\boldsymbol{\mu}\cdot\boldsymbol{p}_i
	}{q_i},
	\qquad
	\boldsymbol{s}_{p,i,k}
	=
	(\boldsymbol{s}_x,s_{p,i,k}).
	\label{eq:pole-locations-snplus1}
\end{equation}
Vectors with \(q_i=0\) do not produce poles in \(s_{n+1}\).

For the small-\(x_{n+1}\) expansion, we consider
only the poles lying to the left of the original convergence strip, according to \cite[Section~2, Theorem~4]{FlajoletGourdonDumas1995}. The conditions for convergence \(\operatorname{Re}\varphi_i(\boldsymbol{s})<0\) give \(\operatorname{Re}s_{n+1}<\operatorname{Re}s_{p,i,0}(\boldsymbol{s}_x)\) if \(q_i>0\); and \(\operatorname{Re}s_{n+1}>\operatorname{Re}s_{p,i,0}(\boldsymbol{s}_x)\) if \(q_i<0\). Therefore to be to the left of the convergence region, we need \(q_i<0\). From now on we consider only such poles whose corresponding \(q_i<0\).

This pole can be of higher order if other facet normal vectors also give a pole at the same value of \(s_{n+1}\).
If \(s_{n+1}=s_{p,i,k}\) is a pole of order \(m\), its contribution
under inverse Mellin transformation is
\begin{align}
	&\underset{s_{n+1}=s_{p,i,k}}{\operatorname{Res}}
	\left[
	t^{-s_{n+1}}\mathcal{M}[I](\boldsymbol{s}_x,s_{n+1})
	\right]
	\notag\\
	&\quad=
	\frac{1}{(m-1)!}
	\left.
	\frac{d^{m-1}}{d s_{n+1}^{m-1}}
	\left[
	(s_{n+1}-s_{p,i,k})^m
	t^{-s_{n+1}}
	\mathcal{M}[I](\boldsymbol{s}_x,s_{n+1})
	\right]
	\right|_{s_{n+1}=s_{p,i,k}}.
	\label{eq:higher-order-pole-inverse-mellin}
\end{align}
In particular, this is \(t^{-s_{p,i,k}}\) times a polynomial in
\(\log t\) of degree at most \(m-1\). The locations of the poles depend
on \(\boldsymbol{s}_x\). We can shift \(\boldsymbol{s}_x\) slightly by some \(\varepsilon\boldsymbol{d}\) to separate the poles. To see this, suppose that the distinct pole-location functions
\(s_{p,1}(\boldsymbol{s}_x),\ldots,s_{p,m}(\boldsymbol{s}_x)\) coincide
at \(\boldsymbol{s}_x=\boldsymbol{s}_{x0}\), with common value
\(s_p^\ast\).
Let
\(\boldsymbol{v}_\alpha=(\boldsymbol{v}_{\alpha,x},q_\alpha)\)
be the normal vector associated with \(s_{p,\alpha}\). From
Eq.~\eqref{eq:pole-locations-snplus1},
\begin{equation}
	s_{p,\alpha}(\boldsymbol{s}_{x0}+\varepsilon\boldsymbol{d})
	=
	s_p^\ast
	-\varepsilon\boldsymbol{d}\cdot
	\frac{\boldsymbol{v}_{\alpha,x}}{q_\alpha}.
	\label{eq:shifted-coalescing-pole-location}
\end{equation}
The normalized vectors
\(\boldsymbol{v}_\alpha/q_\alpha\) belonging to the coalescing poles
are distinct, and hence so are their projections
\(\boldsymbol{v}_{\alpha,x}/q_\alpha\). For every pair
\(\alpha\neq\beta\), the vectors \(\boldsymbol{d}\) satisfying
\begin{equation}
	\boldsymbol{d}\cdot
	\left(
		\frac{\boldsymbol{v}_{\alpha,x}}{q_\alpha}
		-
		\frac{\boldsymbol{v}_{\beta,x}}{q_\beta}
	\right)
	=0
\end{equation}
form a proper hyperplane. A finite union of such hyperplanes cannot
fill \(\mathbb{R}^n\), so we can choose \(\boldsymbol{d}\) outside their
union. For this choice,
\begin{equation}
	s_{p,\alpha}(\boldsymbol{s}_{x0}+\varepsilon\boldsymbol{d})
	\neq
	s_{p,\beta}(\boldsymbol{s}_{x0}+\varepsilon\boldsymbol{d}),
	\qquad
	\alpha\neq\beta,
	\qquad
	0<|\varepsilon|\ll1.
\end{equation}
Therefore nonzero \(\varepsilon\) separates all the \(m\) poles. We can show that taking the limit \(\varepsilon \to 0\) for the sum over all simple separated poles gives the same result as one pole of order \(m\).

Let \(\Gamma\) be a fixed positively oriented contour in the complex plane of \(s_{n+1}\) enclosing these
\(m\) poles and no others for all sufficiently small \(\varepsilon\).
The sum of their simple-pole contributions is
\begin{align}
	&\sum_{\alpha=1}^m
	\underset{
		s_{n+1}=s_{p,\alpha}(
		\boldsymbol{s}_{x0}+\varepsilon\boldsymbol{d})
	}{\operatorname{Res}}
	\left[
		t^{-s_{n+1}}
		\mathcal{M}[I](
		\boldsymbol{s}_{x0}+\varepsilon\boldsymbol{d},s_{n+1})
	\right]
	\notag\\
	&\qquad=
	\frac{1}{2\pi i}
	\oint_\Gamma
	t^{-s_{n+1}}
	\mathcal{M}[I](
	\boldsymbol{s}_{x0}+\varepsilon\boldsymbol{d},s_{n+1})
	ds_{n+1}.
\end{align}
Since the poles stay away from the contour \(\Gamma\), we may therefore take the limit \(\varepsilon\to 0\)
under the contour integral and obtain
\begin{align}
	&\lim_{\varepsilon\to0}
	\sum_{\alpha=1}^m
	\underset{
		s_{n+1}=s_{p,\alpha}(
		\boldsymbol{s}_{x0}+\varepsilon\boldsymbol{d})
	}{\operatorname{Res}}
	\left[
		t^{-s_{n+1}}
		\mathcal{M}[I](
		\boldsymbol{s}_{x0}+\varepsilon\boldsymbol{d},s_{n+1})
	\right]
	\notag\\
	&\qquad=
	\underset{s_{n+1}=s_p^\ast}{\operatorname{Res}}
	\left[
		t^{-s_{n+1}}
		\mathcal{M}[I](\boldsymbol{s}_{x0},s_{n+1})
	\right].
	\label{eq:coalescing-simple-poles-limit}
\end{align}
This is equal to the residue of the order \(m\) pole. Therefore instead of considering higher order poles we may calculate
the simple-pole coefficients for generic \(\boldsymbol{s}_x\) and sum all
contributions from them.

The residues of the separated poles can diverge individually in this
limit; the statement applies to their sum.
For example, suppose that exactly two poles coalesce and set
\begin{equation}
	a_\alpha
	:=
	\boldsymbol{d}\cdot
	\frac{\boldsymbol{v}_{\alpha,x}}{q_\alpha},
	\qquad
	s_{p,\alpha}(\boldsymbol{s}_{x0}+\varepsilon\boldsymbol{d})
	=
	s_p^\ast-\varepsilon a_\alpha,
	\qquad
	a_1\neq a_2.
\end{equation}
Locally, the inverse Mellin integrand then has the form
\begin{equation}
	t^{-s_{n+1}}
	\mathcal{M}[I](
	\boldsymbol{s}_{x0}+\varepsilon\boldsymbol{d},s_{n+1})
	=
	\frac{h(\varepsilon,s_{n+1})}{
		\bigl(s_{n+1}-s_{p,1}\bigr)
		\bigl(s_{n+1}-s_{p,2}\bigr)},
\end{equation}
where \(h\) is holomorphic and
\(h(0,s_p^\ast)\neq0\). The residue at the first pole is
\begin{align}
	\underset{s_{n+1}=s_{p,1}}{\operatorname{Res}}
	\left[
		t^{-s_{n+1}}
		\mathcal{M}[I](
		\boldsymbol{s}_{x0}+\varepsilon\boldsymbol{d},s_{n+1})
	\right]
	&=
	\frac{h(\varepsilon,s_{p,1})}
	{s_{p,1}-s_{p,2}}
	\notag\\
	&=
	\frac{h(\varepsilon,s_{p,1})}
	{\varepsilon(a_2-a_1)}
	=
	\frac{h(0,s_p^\ast)}
	{\varepsilon(a_2-a_1)} + O(1),
	\qquad \varepsilon\to0.
	\label{eq:individual-coalescing-residue-divergence}
\end{align}
Thus this residue diverges as \(1/\varepsilon\); the residue at the
second pole has the opposite leading divergence. Their sum is the
difference quotient
\begin{equation}
	\frac{
		h(\varepsilon,s_{p,1})
		-h(\varepsilon,s_{p,2})
	}{s_{p,1}-s_{p,2}}
	\longrightarrow
	\frac{\partial h}{\partial s_{n+1}}(0,s_p^\ast),
\end{equation}
which is the residue of
\(h(0,s_{n+1})/(s_{n+1}-s_p^\ast)^2\). More generally, suppose that
\(m\) simple poles
\(s_{p,\alpha}=s_p^\ast-\varepsilon a_\alpha\) merge and that locally
the inverse Mellin integrand is
\begin{equation}
	\frac{g(\varepsilon,s_{n+1})}
	{\displaystyle
		\prod_{\beta=1}^m
		\bigl(s_{n+1}-s_{p,\beta}\bigr)},
\end{equation}
with \(g(0,s_p^\ast)\neq0\). The residue at the \(\alpha\)-th pole is
\begin{align}
	\frac{g(\varepsilon,s_{p,\alpha})}
	{\displaystyle
		\prod_{\substack{\beta=1\\\beta\neq\alpha}}^m
		\bigl(s_{p,\alpha}-s_{p,\beta}\bigr)}
	&=
	\frac{g(\varepsilon,s_{p,\alpha})}
	{\displaystyle
		\varepsilon^{m-1}
		\prod_{\substack{\beta=1\\\beta\neq\alpha}}^m
		(a_\beta-a_\alpha)}
	\notag\\
	&=
	O\!\left(\varepsilon^{-(m-1)}\right).
	\label{eq:order-m-coalescing-residue-divergence}
\end{align}
Thus an order-\(m\) pole produces individual separated
residues with poles of order \(m-1\) in \(\varepsilon\); these
divergences cancel only after all \(m\) residues are summed.

\subsubsection{Residues of simple poles}
From now on we will therefore consider only simple poles. We take \(N_i=k+1\) in
Eq.~\eqref{eq:fixed-direction-Ni-integrations-operator}. This gives
\begin{equation}
	\mathcal{M}[I]
	=
	\frac{(-1)^{k+1}J_i}
	{\displaystyle
		\prod_{\ell=0}^{k}
		\bigl(\varphi_i(\boldsymbol{s})-\ell\bigr)}
	\int_{\mathbb{R}^n}d^n\boldsymbol{w}_i\,
	e^{\boldsymbol{s}\cdot\boldsymbol{w}_i}
	\int_{-\infty}^{\infty}d z_i\,
	e^{(\varphi_i(\boldsymbol{s})-k-1)z_i}
	\bigl(e^{z_i}\partial_i\bigr)^{k+1}
	\left[
	\boldsymbol{P}_i^{\boldsymbol{\mu}}
	\right].
	\label{eq:residue-ibp-representation}
\end{equation}
The transverse space \(W_i\) may be chosen freely, provided that
\[
W_i\oplus\operatorname{span}\{\boldsymbol{v}_i\}
=
\mathbb{R}^{n+1}.
\]
Since \(q_i\neq0\), we can choose
\(\boldsymbol{w}_i=(\boldsymbol{y}_x,0)\). Thus
\begin{equation}
	\boldsymbol{y}
	=
	(\boldsymbol{y}_x+z_i\boldsymbol{v}_{i,x},q_iz_i),
	\qquad
	J_i=|q_i|,
	\qquad
	e^{\boldsymbol{s}\cdot\boldsymbol{w}_i}
	=
	e^{\boldsymbol{s}_x\cdot\boldsymbol{y}_x}.
	\label{eq:yx-zi-coordinate-choice}
\end{equation}
The integral becomes
\begin{equation}
	\mathcal{M}[I]
	=
	\frac{(-1)^{k+1}|q_i|}
	{\displaystyle
		\prod_{\ell=0}^{k}
		\bigl(\varphi_i(\boldsymbol{s})-\ell\bigr)}
	\int_{\mathbb{R}^n}d^n\boldsymbol{y}_x\,
	e^{\boldsymbol{s}_x\cdot\boldsymbol{y}_x}
	\int_{-\infty}^{\infty}d z_i\,
	e^{(\varphi_i(\boldsymbol{s})-k)z_i}
	\partial_i
	\left\{
	\bigl(e^{z_i}\partial_i\bigr)^k
	\left[
	\boldsymbol{P}_i^{\boldsymbol{\mu}}
	\right]
	\right\}.
	\label{eq:residue-ibp-yx-representation}
\end{equation}
In the prefactor there is a pole at
\[
\varphi_i(\boldsymbol{s})-k
=
q_i\bigl(s_{n+1}-s_{p,i,k}\bigr)=0.
\]
Its contribution to the series expansion is
\begin{equation}
	t^{-s_{p,i,k}(\boldsymbol{s}_x)}
	A_{i,k}(\boldsymbol{s}_x),
	\qquad
	A_{i,k}(\boldsymbol{s}_x)
	:=
	\underset{s_{n+1}=s_{p,i,k}}{\operatorname{Res}}
	\mathcal{M}[I](\boldsymbol{s}_x,s_{n+1}),
	\label{eq:simple-pole-contribution}
\end{equation}
\begin{equation}
	A_{i,k}(\boldsymbol{s}_x)
	=
	\left.
	\frac{(-1)^k}{k!}
	\int_{\mathbb{R}^n}d^n\boldsymbol{y}_x\,
	e^{\boldsymbol{s}_x\cdot\boldsymbol{y}_x}
	\int_{-\infty}^{\infty}d z_i\,
	e^{(\varphi_i(\boldsymbol{s})-k)z_i}
	\partial_i
	\left\{
	\bigl(e^{z_i}\partial_i\bigr)^k
	\left[
	\boldsymbol{P}_i^{\boldsymbol{\mu}}
	\right]
	\right\}
	\right|_{\varphi_i(\boldsymbol{s})=k}.
	\label{eq:simple-pole-coefficient-integral}
\end{equation}

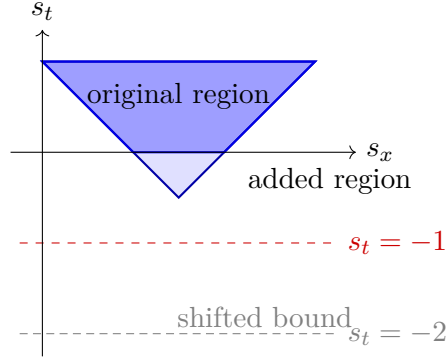
\begin{figure}[ht]
	\centering
	\begin{tikzpicture}[x=1.2cm,y=1.2cm]
		\fill[blue!12]
		(0,1) -- (3,1) -- (1.5,-0.5) -- cycle;
		\fill[blue!38]
		(1,0) -- (2,0) -- (3,1) -- (0,1) -- cycle;
		
		\draw[thick,blue!65!black]
		(0,1) -- (3,1) -- (1.5,-0.5) -- cycle;
		\draw[thick,blue!90!black]
		(1,0) -- (2,0) -- (3,1) -- (0,1) -- cycle;
		
		\draw[dashed,red!80!black]
		(-0.25,-1) -- (3.25,-1)
		node[right] {\(s_t=-1\)};
		\draw[densely dashed,gray]
		(-0.25,-2) -- (3.25,-2)
		node[right] {\(s_t=-2\)};
		
		\draw[->] (-0.35,0) -- (3.45,0)
		node[right] {\(s_x\)};
		\draw[->] (0,-2.25) -- (0,1.35)
		node[above] {\(s_t\)};
		
		\node at (1.5,0.62) {original region};
		\node[anchor=west] at (2.15,-0.32) {added region};
		\node[gray] at (2.45,-1.82) {shifted bound};
	\end{tikzpicture}
	\caption{The dark trapezoid
		is the original convergence region, and the outlined light triangle is the added region after
		integrating by parts two times. The pole \(s_t=-1\) remains
		outside because the slanted inequalities meet at \(s_t=-1/2\), even
		though the shifted directional bound is \(s_t>-2\).}
	\label{fig:two-dimensional-convergence-regions}
\end{figure}

However, the inequality \(\operatorname{Re}\varphi_i(\boldsymbol{s})<k+1\) alone does not ensure
that the integral for \(A_{i,k}\)
converges at \(\varphi_i(\boldsymbol{s})=k\), because the convergence
conditions in the other directions must still hold. For a simple
two-dimensional example with real \(s_x\) and \(s_t\), consider
\begin{equation}
	\mathcal{M}(s_x,s_t)
	=
	\int_0^\infty\frac{dx}{x}
	\int_0^\infty\frac{dt}{t}\,
	\frac{x^{s_x}t^{s_t}}
	{(t+x+x^2+x^3t)}.
	\label{eq:two-dimensional-convergence-example}
\end{equation}
Integrating by parts two times using \(\boldsymbol{v}_i=(0,-1)\), the lower bound moves from \(s_t>0\) to \(s_t>-2\). Figure~\ref{fig:two-dimensional-convergence-regions}
shows the full convergence regions after the conditions from the other
facets are also imposed. These other facets already stop
the enlarged region at \(s_t=-1/2\).

For \(k=1\), the pole is at \(s_t=-1\). It satisfies the shifted
directional inequality \(s_t>-2\), but Figure~\ref{fig:two-dimensional-convergence-regions}
shows that the pole still lies outside the convergence region of the integral.

Therefore, in general we cannot set \(\varphi_i(\boldsymbol{s})=k\) directly in Eq.~\eqref{eq:simple-pole-coefficient-integral} and calculate the meromorphic continuation in \(\boldsymbol{s}_{x}\), because as seen from the example the remaining integral might not converge for any value of \(\boldsymbol{s}_{x}\).
Instead we should have started with an equation of the form Eq.~\eqref{eq:all-direction-shifted-operator}, where there have been more integrations by parts done in other facet normal directions, such that the integral converges when setting \(\varphi_i(\boldsymbol{s})=k\), and used that to calculate the residue. That is equal to the unique meromorphic continuation of Eq.~\eqref{eq:simple-pole-coefficient-integral} evaluated at \(\varphi_i(\boldsymbol{s})=k\).

\subsubsection{Meromorphic continuation per sector}
However, we can meromorphically continue Eq.~\eqref{eq:simple-pole-coefficient-integral} in a different way. Note that \(\varphi_i(\boldsymbol{s})\) depends on \(s_{n+1}\) and by changing \(s_{n+1}\) we can set \(\varphi_i(\boldsymbol{s})\) to any desired value, so we can view \(\varphi_i\) as an independent variable (\(\boldsymbol{s}_{x}\) and \(\varphi_i\) are sometimes called regulators). Define the integral \(A_{i,k}(\boldsymbol{s}_x,\varphi_i)\), from which the residue is \(A_{i,k}(\boldsymbol{s}_x)=A_{i,k}(\boldsymbol{s}_x,k)\). For values of \(\varphi_i\) and \(\boldsymbol{s}_{x}\) where the integral converges, we can subdivide the transverse space
\(\boldsymbol{y}_x\in\mathbb{R}^n\) into \(n\)-dimensional sectors \(\{C_{\rho,x}\}\). Then the integral is given by the sum over all sectors:
\begin{align}
	A_{i,k}(\boldsymbol{s}_x,\varphi_i)
	={}&
	\sum_\rho
	A_{i,k}^{(\rho)}(\boldsymbol{s}_x,\varphi_i)
	\\
	A_{i,k}^{(\rho)}(\boldsymbol{s}_x,\varphi_i)
	:={}&
	\frac{(-1)^k}{k!}
	\int_{C_{\rho,x}}d^n\boldsymbol{y}_x\,
	e^{\boldsymbol{s}_x\cdot\boldsymbol{y}_x}
	\int_{-\infty}^{\infty}d z_i\,
	e^{(\varphi_i-k)z_i}
	\partial_i
	\left\{
		\bigl(e^{z_i}\partial_i\bigr)^k
		\left[
			\boldsymbol{P}_i^{\boldsymbol{\mu}}
		\right]
	\right\}.
	\label{eq:one-sector-residue-integral}
\end{align}
Since the integral over the whole space converges absolutely for some range of the regulators, also each sector integral converges for at least the same range. 
Because the integral might not be directly calculable at the final values of \(\boldsymbol{s}_{x}\) and \(\varphi_i=k\) that we are interested in, we have to calculate the meromorphic continuation of it (as we discussed according to the theory of Mellin transforms the series expansion we get from the poles and residues of the meromorphic continuation of the Mellin transform).

If each sector integral has a joint meromorphic continuation (\(\operatorname{MC}\)) in the regulators, then in general since the sum of meromorphic continuations is the meromorphic continuation of the sum, we can write:
\begin{equation}
	\operatorname{MC}_{\boldsymbol{s}_x,\varphi_i}
	\left\{
	A_{i,k}(\boldsymbol{s}_x,\varphi_i)
	\right\}
	=
	\sum_\rho
	\operatorname{MC}_{\boldsymbol{s}_x,\varphi_i}
	\left\{
	A_{i,k}^{(\rho)}(\boldsymbol{s}_x,\varphi_i)
	\right\}.
\end{equation}

Furthermore, if each sector integral is defined for some values of \(\boldsymbol{s}_{x}\) along the line \(\varphi_i=k\), we can immediately insert \(\varphi_i=k\) into the integral and do meromorphic continuation only in \(\boldsymbol{s}_{x}\), because a joint meromorphic continuation is unique if it exists and we can continue to the final point through a path we choose. This gives for one sector integral from Eq.~\eqref{eq:one-sector-residue-integral}:
\begin{align}
	A_{i,k}^{(\rho)}(\boldsymbol{s}_x,k)
	={}&
	\frac{(-1)^k}{k!}
	\int_{C_{\rho,x}}d^n\boldsymbol{y}_x\,
	e^{\boldsymbol{s}_x\cdot\boldsymbol{y}_x}
	\int_{-\infty}^{\infty}d z_i\,
	\partial_i
	\left\{
		\bigl(e^{z_i}\partial_i\bigr)^k
		\left[
			\boldsymbol{P}_i^{\boldsymbol{\mu}}
		\right]
	\right\}
	\notag\\
	={}&
	\frac{(-1)^k}{k!}
	\int_{C_{\rho,x}}d^n\boldsymbol{y}_x\,
	e^{\boldsymbol{s}_x\cdot\boldsymbol{y}_x}
	\left[
		\bigl(e^{z_i}\partial_i\bigr)^k
		\left(
			\boldsymbol{P}_i^{\boldsymbol{\mu}}
		\right)
	\right]_{z_i=-\infty}^{z_i=+\infty}.
\end{align}
To see why the limit \(z_i=-\infty\) is zero,  we know
from Eqs.~\eqref{eq:fixed-direction-Ni-integrations-operator}--
\eqref{eq:fixed-direction-Ni-integrations} and
\eqref{eq:fixed-direction-tilde-T-scaling-replacement} that
\begin{equation}
	\bigl(e^{z_i}\partial_i\bigr)^k
	\left[
		\boldsymbol{P}_i^{\boldsymbol{\mu}}
	\right]
	=
	\boldsymbol{P}_i^{\boldsymbol{\mu}}
	\frac{\widetilde{T}_{(i,k)}}
	{\boldsymbol{P}_i^{k\boldsymbol{1}}}
	\cong
	\boldsymbol{P}_i^{\boldsymbol{\mu}}
	\frac{
		\bigl(e^{z_i}\partial_i\bigr)^k
		\left[
		\boldsymbol{P}_i^{k\boldsymbol{1}}
		\right]}
	{\boldsymbol{P}_i^{k\boldsymbol{1}}}.
	\label{eq:kfold-tilde-T-scaling-replacement}
\end{equation}
Let the largest power of each \(P_\ell\) in the
\(-\boldsymbol{v}_i\)-direction be
\begin{equation}
	p'_{i,\ell}
	:=
	\max_{1\leq a\leq N_\ell}
	\boldsymbol{a}_{\ell,a}\cdot(-\boldsymbol{v}_i),
	\qquad
	\boldsymbol{p}'_i
	:=
	(p'_{i,1},\ldots,p'_{i,M}).
	\label{eq:largest-opposite-directional-powers}
\end{equation}
The largest \(e^{-z_i}\)-power of
\(\boldsymbol{P}_i^{k\boldsymbol{1}} = \exp(-k\boldsymbol{1}\cdot\boldsymbol{p}_iz_i) \boldsymbol{P}^{k\boldsymbol{1}}\) is
\(k\boldsymbol{1}\cdot(\boldsymbol{p}'_i+\boldsymbol{p}_i)\). Each application of
\(e^{z_i}\partial_i\) reduces the \(e^{-z_i}\)-power of every surviving
monomial by one, so the operator
\(\bigl(e^{z_i}\partial_i\bigr)^k\) reduces it by \(k\). Consequently,
the largest \(e^{-z_i}\)-power that can occur in
\(\widetilde{T}_{(i,k)}\) is
\begin{align}
	&k\boldsymbol{1}\cdot
	(\boldsymbol{p}'_i+\boldsymbol{p}_i)
	-k
	\notag\\
	&\qquad=
	k\left(
	\boldsymbol{1}\cdot
	(\boldsymbol{p}'_i+\boldsymbol{p}_i)-1
	\right).
	\label{eq:general-term-opposite-directional-power}
\end{align}
Therefore the largest \(e^{-z_i}\)-power of
\begin{equation}
\boldsymbol{P}_i^{\boldsymbol{\mu}}
	\frac{
		\widetilde{T}_{(i,k)}}
	{\boldsymbol{P}_i^{k\boldsymbol{1}}}.
	\label{eq:PmuToverPkfactor}
\end{equation}
is at most \(\boldsymbol{\mu}\cdot\boldsymbol{p}_i+\boldsymbol{\mu}\cdot\boldsymbol{p}'_i-k\). We already know from the text after \eqref{eq:fixed-direction-tilde-T-operator-count} that the largest power of \(e^{z_i}\) of \(\widetilde{T}_{(i,k)}\), and therefore of the whole expression \eqref{eq:PmuToverPkfactor}, is at most zero.
Thus, for fixed \(\boldsymbol{y}_x\), considering the scaling we have just discussed, we can find a finite constant
\(C_{i,k}(\boldsymbol{y}_x)\), independent of \(z_i\), such that
\begin{align}
	\left|
	e^{(\varphi_i(\boldsymbol{s})-k)z_i}
	\bigl(e^{z_i}\partial_i\bigr)^k
	\left(
		\boldsymbol{P}_i^{\boldsymbol{\mu}}
	\right)
	\right|
	&=
	\left|
	e^{(\varphi_i(\boldsymbol{s})-k)z_i}
	\boldsymbol{P}_i^{\boldsymbol{\mu}}
	\frac{\widetilde{T}_{(i,k)}}
	{\boldsymbol{P}_i^{k\boldsymbol{1}}}
	\right|
	\notag\\
	&\leq
	C_{i,k}(\boldsymbol{y}_x)
	\begin{cases}
		\exp\!\left[
		\bigl(\operatorname{Re}\varphi_i(\boldsymbol{s})-k\bigr)z_i
		\right],
		& z_i\geq0,\\[2mm]
		\exp\!\left[
		-\operatorname{Re}\varphi'_i(\boldsymbol{s})z_i
		\right],
		& z_i\leq0,
	\end{cases}
	\label{eq:ratio-operator-two-sided-exponential-scaling}
\end{align}
where
\begin{equation}
	\varphi'_i(\boldsymbol{s})
	:=
	\boldsymbol{s}\cdot(-\boldsymbol{v}_i)
	+\boldsymbol{\mu}\cdot\boldsymbol{p}'_i.
	\label{eq:opposite-direction-scaling-function}
\end{equation}
In the convergence region of the original Mellin transform,
\begin{equation}
	\operatorname{Re}\varphi_i(\boldsymbol{s})<0,
	\qquad
	\operatorname{Re}\varphi'_i(\boldsymbol{s})<0,
\end{equation}
and hence
\begin{equation}
	\operatorname{Re}\varphi_i(\boldsymbol{s})
	+\operatorname{Re}\varphi'_i(\boldsymbol{s})
	=
	\boldsymbol{\mu}\cdot
	(\boldsymbol{p}_i+\boldsymbol{p}'_i)
	<0.
	\label{eq:opposite-direction-scaling-sum}
\end{equation}
This sum is independent of \(\boldsymbol{s}\). Therefore, if we set \(s_{n+1}\) such that
\(\varphi_i(\boldsymbol{s})=k\), we have
\(\operatorname{Re}\varphi'_i(\boldsymbol{s})<-k\). From this \(\exp\!\left[
		-\operatorname{Re}\varphi'_i(\boldsymbol{s})z_i
		\right]\)
tends to zero as \(z_i\to-\infty\). Hence only the boundary at \(z_i=+\infty\) remains, and
the contribution from one sector becomes
\begin{equation}
	A_{i,k}^{(\rho)}(\boldsymbol{s}_x,k)
	=
	\frac{(-1)^k}{k!}
	\int_{C_{\rho,x}}d^n\boldsymbol{y}_x\,
	e^{\boldsymbol{s}_x\cdot\boldsymbol{y}_x}
	\lim_{z_i\to+\infty}
	\bigl(e^{z_i}\partial_i\bigr)^k
	\left(
		\boldsymbol{P}_i^{\boldsymbol{\mu}}
	\right).
\end{equation}
In summary we conclude that if for each sector a meromorphic continuation of \(A_{i,k}^{(\rho)}(\boldsymbol{s}_x,\varphi_i)\) exists in all the regulators and if a meromorphic continuation of \(A_{i,k}^{(\rho)}(\boldsymbol{s}_x,k)\) exists in \(\boldsymbol{s}_x\), then summing \(A_{i,k}^{(\rho)}(\boldsymbol{s}_x,k)\) together over all sectors gives us the desired residue \(A_{i,k}\).

\subsubsection{Cones as sectors}

Now we will pick cones \(\{C_{\rho,x}\}\) (each cone has \(n\) rational linearly independent generators) as sectors in the \(\boldsymbol{y}_x\) space
(for example sector decomposition \cite{Heinrich2008} is also equivalent to picking such cones as sectors) and show that in that case the sector integrals have meromorphic continuations. We can show it using an argument similar to how we showed that Eq.~\eqref{eq:all-direction-shifted-operator} is a meromorphic continuation of \(\mathcal{M}[I]\). To see this, we take one sector integral
\begin{align}
	A_{i,k}^{(\rho)}
	\bigl(\boldsymbol{s}_{x},\varphi_{i}\bigr)
	={}&
	\frac{(-1)^k}{k!}
	\int_{C_{\rho,x}}d^n\boldsymbol{y}_x\,
	e^{\boldsymbol{s}_{x}\cdot\boldsymbol{y}_x}
	\int_{-\infty}^{\infty}d z_i\,
	e^{(\varphi_{i}-k)z_i}
	\partial_i
	\left\{
		\bigl(e^{z_i}\partial_i\bigr)^k
		\left[
			\boldsymbol{P}_i^{\boldsymbol{\mu}}
		\right]
	\right\}.
\end{align}
and write
\begin{equation}
	\partial_i
	\left\{
		\bigl(e^{z_i}\partial_i\bigr)^k
		\left[
			\boldsymbol{P}_i^{\boldsymbol{\mu}}
		\right]
	\right\}
	=
	e^{-z_i}
	\boldsymbol{P}_i^{\boldsymbol{\mu}}
	\frac{\widetilde{T}_{(i,k+1)}}
	{\boldsymbol{P}_i^{(k+1)\boldsymbol{1}}},
	\label{eq:simple-pole-derivative-PT-form}
\end{equation}
where we used
Eq.~\eqref{eq:fixed-direction-tilde-T-operator-action}. Write \(\widetilde{T}_{(i,k+1)}\) as a finite sum of monomials:
\begin{equation}
	\widetilde{T}_{(i,k+1)}
	=
	\sum_{\alpha\in E_{i,k+1}}
	c_\alpha
	e^{\boldsymbol{a}_{T,\alpha}\cdot
	(\boldsymbol{y}_x,z_i)}.
\end{equation}
It is enough to show that we can meromorphically continue the integral considering only one monomial and then sum all the meromorphic continuations together.

We can subdivide the integration space \(C_{\rho,x}\times\mathbb{R}\) into simplicial cones where a specific vertex of \(\Delta_{\boldsymbol{P}}\) has maximal power. Let
\(C_\sigma\) be one of such
\((n+1)\)-dimensional cones.
Let its linearly independent integer generators be
\(\boldsymbol{w}_{\sigma,1},\ldots,\boldsymbol{w}_{\sigma,n+1}\).  We use
\begin{equation}
	(\boldsymbol{y}_x,z_i)
	=
	\sum_{b=1}^{n+1}
	\xi_{\sigma,b}\boldsymbol{w}_{\sigma,b},
	\qquad
	\xi_{\sigma,b}\geq0,
	\qquad
	d^n\boldsymbol{y}_x\,dz_i
	=
	J_\sigma\prod_{b=1}^{n+1}d\xi_{\sigma,b}.
	\label{eq:refined-full-cone-coordinates}
\end{equation}
On \(C_\sigma\), a fixed monomial of each \(P_{i,\ell}\) is dominant.
If its exponent vector is
\(\boldsymbol{a}_{\sigma,\ell}\), set
\begin{equation}
	p_{\sigma,b,\ell}
	:=
	\boldsymbol{a}_{\sigma,\ell}\cdot\boldsymbol{w}_{\sigma,b},
	\qquad
	\boldsymbol{p}_{\sigma,b}
	:=
	(p_{\sigma,b,1},\ldots,p_{\sigma,b,M}),
\end{equation}
and factor
\begin{equation}
	P_{i,\ell}
	=
	e^{\sum_b p_{\sigma,b,\ell}\xi_{\sigma,b}}
	\widehat P_{\sigma,\ell},
	\qquad \ell=1,\ldots,M.
\end{equation}
The remaining exponential powers of \(e^{\xi_{\sigma,b}}\) in
\(\widehat P_{\sigma,\ell}\) are nonpositive for all \(b\). The contribution of the monomial
\(\alpha\) of \(\widetilde{T}_{(i,k+1)}\) on \(C_\sigma\) is therefore
\begin{equation}
	\frac{(-1)^kc_\alpha J_\sigma}{k!}
	\int_{[0,\infty)^{n+1}}
	\prod_{b=1}^{n+1}d\xi_{\sigma,b}\,
	e^{\sum_{b=1}^{n+1}
	\lambda_{\sigma,b}\xi_{\sigma,b}}
	\widehat{\boldsymbol{P}}_\sigma^{
		\boldsymbol{\mu}-(k+1)\boldsymbol{1}},
	\label{eq:refined-simplicial-cone-monomial-integral}
\end{equation}
where
\(\widehat{\boldsymbol{P}}_\sigma
:=(\widehat P_{\sigma,1},\ldots,\widehat P_{\sigma,M})\) and
\begin{align}
	\lambda_{\sigma,b}
	={}&
	\bigl(\boldsymbol{s}_{x},
	\varphi_{i}-k-1\bigr)
	\cdot\boldsymbol{w}_{\sigma,b}
	+
	\boldsymbol{a}_{T,\alpha}\cdot\boldsymbol{w}_{\sigma,b}
	\notag\\
	&+
	\bigl(\boldsymbol{\mu}-(k+1)\boldsymbol{1}\bigr)
	\cdot\boldsymbol{p}_{\sigma,b}.
\end{align}
Fix \(b\). In the initial convergence region, integration by parts in
\(\xi_{\sigma,b}\) gives
\begin{align}
	&\frac{(-1)^kc_\alpha J_\sigma}{k!}
	\int_{[0,\infty)^{n+1}}
	\prod_{c=1}^{n+1}d\xi_{\sigma,c}\,
	e^{\sum_{c=1}^{n+1}
		\lambda_{\sigma,c}\xi_{\sigma,c}}
	\widehat{\boldsymbol{P}}_\sigma^{
		\boldsymbol{\mu}-(k+1)\boldsymbol{1}}
	\notag\\
	&\quad=
	-\frac{(-1)^kc_\alpha J_\sigma}
	{k!\lambda_{\sigma,b}}
	\int_{[0,\infty)^n}
	\prod_{\substack{c=1\\c\neq b}}^{n+1}
	d\xi_{\sigma,c}\,
	\left.
	e^{\sum_{\substack{c=1}}^{n+1}
		\lambda_{\sigma,c}\xi_{\sigma,c}}
	\widehat{\boldsymbol{P}}_\sigma^{
		\boldsymbol{\mu}-(k+1)\boldsymbol{1}}
	\right|_{\xi_{\sigma,b}=0}
	\label{eq:refined-cone-coordinate-ibp1}\\
	&\qquad
	-\frac{(-1)^kc_\alpha J_\sigma}
	{k!\lambda_{\sigma,b}}
	\int_{[0,\infty)^{n+1}}
	\prod_{c=1}^{n+1}d\xi_{\sigma,c}\,
	e^{\sum_{c=1}^{n+1}
		\lambda_{\sigma,c}\xi_{\sigma,c}}
	\partial_{\xi_{\sigma,b}}
	\left[
	\widehat{\boldsymbol{P}}_\sigma^{
		\boldsymbol{\mu}-(k+1)\boldsymbol{1}}
	\right].
	\label{eq:refined-cone-coordinate-ibp}
\end{align}
The derivative in the last integral is
\begin{align}
	&\partial_{\xi_{\sigma,b}}
	\left[
	\widehat{\boldsymbol{P}}_\sigma^{
		\boldsymbol{\mu}-(k+1)\boldsymbol{1}}
	\right]
	\notag\\
	&\quad=
	\widehat{\boldsymbol{P}}_\sigma^{
		\boldsymbol{\mu}-(k+2)\boldsymbol{1}}
	U_{\sigma,b},
	\label{eq:refined-cone-P-derivative}
\end{align}
where
\begin{equation}
	U_{\sigma,b}
	:=
	\sum_{\ell=1}^M
	(\mu_\ell-k-1)
	\left(
	\partial_{\xi_{\sigma,b}}\widehat P_{\sigma,\ell}
	\right)
	\prod_{\substack{h=1\\h\neq\ell}}^M
	\widehat P_{\sigma,h}.
\end{equation}
Each \(\widehat P_{\sigma,\ell}\) contains only nonpositive exponential
powers in every \(\xi_{\sigma,c}\). The derivative removes the monomials
with zero \(\xi_{\sigma,b}\)-power. Therefore \(U_{\sigma,b}\) is a finite
sum
\begin{equation}
	U_{\sigma,b}
	=
	\sum_{\eta\in E_{\sigma,b}}
	d_\eta
	\exp\left(
	\sum_{c=1}^{n+1}
	u_{\eta,c}\xi_{\sigma,c}
	\right),
	\qquad
	u_{\eta,c}\leq0,
	\qquad
	u_{\eta,b}\leq-1.
\end{equation}
The factor
\(\widehat{\boldsymbol{P}}_\sigma^{
\boldsymbol{\mu}-(k+2)\boldsymbol{1}}\)
is bounded on \(C_\sigma\), so it does not
change the exponential scaling.
For each monomial in this sum, the maximal exponential powers of the second
integral from integration by parts (Eq.~\eqref{eq:refined-cone-coordinate-ibp}) become
\(\lambda_{\sigma,b}+u_{\eta,b}\). Thus the power in
\(\xi_{\sigma,b}\) decreases by at least one, while the powers in the
other cone coordinates cannot increase and may also decrease. Before integration by parts the condition for that integral to converge considering the \(\boldsymbol{w}_{\sigma,b}\) direction was \(\operatorname{Re}\lambda_{\sigma,b}<0\), then now it is \(\operatorname{Re}\lambda_{\sigma,b}<1\) at least. We can
then repeat the same integration-by-parts step for each monomial of \(U_{\sigma,b}\), either
in the same coordinate or in another cone coordinate. Applying the
same procedure repeatedly also to the first integral \eqref{eq:refined-cone-coordinate-ibp1} from integration by parts we can make the integral converge for arbitrary values of \(\boldsymbol{\lambda}_{\sigma}\) and therefore for arbitrary values of \(\boldsymbol{s}_{x}\) and \(\varphi_{i}\), if we do integration by parts enough times. Therefore we can get a
meromorphic continuation to arbitrary values of
\(\boldsymbol{s}_{x}\) and \(\varphi_{i}\).

\subsubsection{Setting \texorpdfstring{\(\varphi_i=k\)}{phi.i = k}}
Next we will show that for the cone \(C_{\rho,x}\) there exist values of \(\boldsymbol{s}_x\) for which the integral converges at \(\varphi_i=k\). Take the \(C_{\rho,x}\) cone's linearly independent generators with rational components as
\(\boldsymbol{w}_{\rho,1},\ldots,
\boldsymbol{w}_{\rho,n}\). We parameterize the cone by
\begin{equation}
	\boldsymbol{y}_x
	=
	\sum_{a=1}^n \xi_{\rho,a}\boldsymbol{w}_{\rho,a},
	\qquad
	\xi_{\rho,a}\geq 0,
	\qquad
	d^n\boldsymbol{y}_x
	=
	J_{\rho,x}\prod_{a=1}^n d\xi_{\rho,a},
	\qquad
	J_{\rho,x}
	=
	\left|
	\det\bigl(
	\boldsymbol{w}_{\rho,1},\ldots,
	\boldsymbol{w}_{\rho,n}
	\bigr)
	\right|.
	\label{eq:transverse-cone-positive-coordinates}
\end{equation}
Let
\(\boldsymbol{w}_{\rho}^{*,1},\ldots,
\boldsymbol{w}_{\rho}^{*,n}\)
be the dual basis, so that
\begin{equation}
	\boldsymbol{w}_{\rho}^{*,a}\cdot
	\boldsymbol{w}_{\rho,b}
	=
	\delta^a_b.
\end{equation}
We write the cone regulator in this dual basis as
\begin{equation}
	\boldsymbol{s}_{x}
	=
	\sum_{a=1}^n
	s_{\rho,a}\boldsymbol{w}_{\rho}^{*,a}.
	\label{eq:transverse-cone-dual-components}
\end{equation}
Introduce the coordinate vectors and measure
\begin{equation}
	\boldsymbol{\xi}_{\rho}
	=
	(\xi_{\rho,1},\ldots,\xi_{\rho,n}),
	\qquad
	\boldsymbol{s}_{\rho}
	=
	(s_{\rho,1},\ldots,s_{\rho,n}),
	\qquad
	d^n\boldsymbol{\xi}_{\rho}
	:=
	\prod_{a=1}^n d\xi_{\rho,a}.
\end{equation}
Then
\(\boldsymbol{s}_{x}\cdot\boldsymbol{y}_x
=\boldsymbol{s}_{\rho}\cdot\boldsymbol{\xi}_{\rho}\).
The cone contribution is
\begin{equation}
	A_{i,k}^{(\rho)}\bigl(\boldsymbol{s}_{x},\varphi_{i}\bigr)=\frac{(-1)^kJ_{\rho,x}}{k!}\int_{[0,\infty)^n}d^n\boldsymbol{\xi}_{\rho}\,e^{\boldsymbol{s}_{\rho}\cdot\boldsymbol{\xi}_{\rho}}\int_{-\infty}^{\infty}d z_i\,e^{(\varphi_{i}-k)z_i}\partial_i\left\{\bigl(e^{z_i}\partial_i\bigr)^k\left[\boldsymbol{P}_i^{\boldsymbol{\mu}}\right]\right\}.
	\label{eq:single-cone-positive-coordinate-integral}
\end{equation}

Inserting \(\varphi_{i}=k\) into
Eq.~\eqref{eq:single-cone-positive-coordinate-integral} to check if it converges for some values of \(\boldsymbol{s}_{\rho}\), we can integrate
the total derivative in \(z_i\) to obtain
\begin{align}
	A_{i,k}^{(\rho)}
	\bigl(\boldsymbol{s}_{x},k\bigr)
	={}&
	\frac{(-1)^kJ_{\rho,x}}{k!}
	\int_{[0,\infty)^n}
	d^n\boldsymbol{\xi}_{\rho}\,
	e^{\boldsymbol{s}_{\rho}\cdot\boldsymbol{\xi}_{\rho}}
	\left[
		\bigl(e^{z_i}\partial_i\bigr)^k
		\left(
		\boldsymbol{P}_i^{\boldsymbol{\mu}}
		\right)
	\right]_{z_i=-\infty}^{z_i=+\infty}.
	\label{eq:single-cone-after-zi-integration}
\end{align}

\begin{align}
	A_{i,k}^{(\rho)}
	\bigl(\boldsymbol{s}_{x},k\bigr)
	={}&
	\frac{(-1)^kJ_{\rho,x}}{k!}
	\int_{[0,\infty)^n}
	d^n\boldsymbol{\xi}_{\rho}\,
	e^{\boldsymbol{s}_{\rho}\cdot\boldsymbol{\xi}_{\rho}}
	\lim_{z_i\to+\infty}
	\bigl(e^{z_i}\partial_i\bigr)^k
	\left(
		\boldsymbol{P}_i^{\boldsymbol{\mu}}
	\right).
	\label{eq:single-cone-positive-zi-boundary}
\end{align}
The lower boundary in \(z_i\) is zero according to the argument after
Eq.~\eqref{eq:opposite-direction-scaling-sum}.
This integral converges if we choose all \(\operatorname{Re}\boldsymbol{s}_\rho\) sufficiently negative, because the integration space is the positive orthant and if all \(\operatorname{Re}\boldsymbol{s}_\rho\) are negative they reduce the overall scaling in all directions of the integration space.

In summary, if we divide the \(\boldsymbol{y}_x\) integration space into any \(n\) dimensional cones, whose \(n\) linearly independent generators have rational components, then each cone integral has a meromorphic continuation, which we can do along the line \(\varphi_i=k\) and therefore the residue that determines the coefficient in the series expansion in \(t\) of the original integral \eqref{eq:introduction-integral} is:
\begin{align}
	A_{i,k}(\boldsymbol{s}_x)
	={}&
	\left.
	\operatorname{MC}_{\boldsymbol{s}_x,\varphi_i}
	\left\{
	\frac{(-1)^k}{k!}
	\int_{\mathbb{R}^n}d^n\boldsymbol{y}_x\,
	e^{\boldsymbol{s}_x\cdot\boldsymbol{y}_x}
	\int_{-\infty}^{\infty}d z_i\,
	e^{(\varphi_i-k)z_i}
	\partial_i
	\left\{
		\bigl(e^{z_i}\partial_i\bigr)^k
		\left[
			\boldsymbol{P}_i^{\boldsymbol{\mu}}
		\right]
	\right\}
	\right\}
	\right|_{
		\varphi_i=k}
	\notag\\
	={}&
	\sum_\rho
	\operatorname{MC}_{\boldsymbol{s}_{x}}
	\left\{
	\frac{(-1)^k}{k!}
	\int_{C_{\rho,x}}d^n\boldsymbol{y}_x\,
	e^{\boldsymbol{s}_{x}\cdot\boldsymbol{y}_x}
	\lim_{z_i\to+\infty}
	\bigl(e^{z_i}\partial_i\bigr)^k
	\left(
		\boldsymbol{P}_i^{\boldsymbol{\mu}}
	\right)
	\right\}.
	\label{eq:pole-sum-cones}
\end{align}

\subsubsection{Residue in \texorpdfstring{\(\boldsymbol{x}\)}{x} space using \texorpdfstring{\(\lambda\)}{lambda}}

For the formulas below, we suppress the explicit sum over sectors \(\rho\),
writing the sector sum as a single integral over \(\mathbb{R}^n\).
This notation does not assert that the integral converges and if it does not, we divide it into
the sectors \(C_{\rho,x}\), then meromorphically continue each result in
\(\boldsymbol{s}_{x}\) to the desired value and sum all sectors together.
Now set
\begin{equation}
	z_i=-\ln r_i.
\end{equation}
As \(z_i\to+\infty\), we have \(r_i\to0^+\), and
\begin{equation}
	e^{z_i}\partial_i
	=
	-\frac{\partial}{\partial r_i}.
	\label{eq:zi-to-ri-operator}
\end{equation}
In these variables, the factorized polynomials are
\begin{equation}
	P_{i,\ell}
	=
	r_i^{p_{i,\ell}}
	P_\ell\!\left(
		e^{\boldsymbol{y}_x}r_i^{-\boldsymbol{v}_{i,x}},
		r_i^{-q_i}
	\right),
	\qquad
	\ell=1,\ldots,M.
	\label{eq:factorized-polynomials-ri}
\end{equation}
Consequently, the boundary in
Eq.~\eqref{eq:pole-sum-cones} is
\begin{equation}
	\lim_{r_i\to0^+}
	\left(-\frac{\partial}{\partial r_i}\right)^k
	\left[
		r_i^{\boldsymbol{\mu}\cdot\boldsymbol{p}_i}
		\boldsymbol{P}\!\left(
			e^{\boldsymbol{y}_x}r_i^{-\boldsymbol{v}_{i,x}},
			r_i^{-q_i}
		\right)^{\boldsymbol{\mu}}
	\right].
	\label{eq:residue-boundary-ri}
\end{equation}
The operator contributes a factor \((-1)^k\), which cancels the
factor \((-1)^k\) in
Eq.~\eqref{eq:pole-sum-cones}, and the residue becomes
\begin{equation}
	A_{i,k}(\boldsymbol{s}_x)
	=
	\frac{1}{k!}
	\int_{\mathbb{R}^n}d^n\boldsymbol{y}_x\,
	e^{\boldsymbol{s}_x\cdot\boldsymbol{y}_x}
	\lim_{r_i\to0^+}
	\left(\frac{\partial}{\partial r_i}\right)^k
	\left[
		r_i^{\boldsymbol{\mu}\cdot\boldsymbol{p}_i}
		\boldsymbol{P}\!\left(
			e^{\boldsymbol{y}_x}r_i^{-\boldsymbol{v}_{i,x}},
			r_i^{-q_i}
		\right)^{\boldsymbol{\mu}}
	\right].
	\label{eq:residue-ri-region-form}
\end{equation}
Under inverse Mellin transformation, this residue is the coefficient
of
\begin{equation}
	t^{-s_{p,i,k}(\boldsymbol{s}_x)}
	=
	t^{\frac{
			\boldsymbol{s}_x\cdot\boldsymbol{v}_{i,x}
			+\boldsymbol{\mu}\cdot\boldsymbol{p}_i-k
		}{q_i}}
	\label{eq:residue-corresponding-t-power}
\end{equation}
in the small-\(t\) expansion.

To use the inward-pointing normal-vector convention of the expansion by
regions, define
\begin{equation}
	\boldsymbol{u}_i
	:=
	\frac{\boldsymbol{v}_i}{q_i}
	=
	(\boldsymbol{\gamma}_i,1).
	\label{eq:inward-normal-and-region-parameter}
\end{equation}
In general, a derivative at the origin may be written using a rescaled
argument as
\begin{align}
	\left.
	\left(\frac{\partial}{\partial r_i}\right)^k
	f(r_i)
	\right|_{r_i=0}
	&=
	\left.
	\left(
	\frac{\partial}{\partial(\lambda t)^{1/(-q_i)}}
	\right)^k
	f\!\left((\lambda t)^{1/(-q_i)}\right)
	\right|_{\lambda=0}
	\notag\\
	&=
	\left.
	\left[
	\frac{-q_i}{t}
	(\lambda t)^{1+1/q_i}
	\frac{\partial}{\partial\lambda}
	\right]^k
	f\!\left((\lambda t)^{1/(-q_i)}\right)
	\right|_{\lambda=0}.
	\label{eq:derivative-rescaling-identity}
\end{align}
Thus we introduce the expansion parameter \(\lambda\) through
\begin{equation}
	r_i=(\lambda t)^{1/(-q_i)},
	\qquad\text{equivalently}\qquad
	\lambda=\frac{r_i^{-q_i}}{t},
\end{equation}
and make this substitution directly under the derivative. Then the complete simple-pole contribution to the series expansion is
\begin{align}
	&t^{-s_{p,i,k}(\boldsymbol{s}_x)}
	A_{i,k}(\boldsymbol{s}_x)
	\notag\\
	&\quad=
	\frac{t^{\boldsymbol{s}_x\cdot\boldsymbol{\gamma}_i}}{k!}
	\int_{\mathbb{R}^n}d^n\boldsymbol{y}_x\,
	e^{\boldsymbol{s}_x\cdot\boldsymbol{y}_x}
	\left.
	\left(
	\frac{\partial}{\partial\lambda^{1/(-q_i)}}
	\right)^k
	\left[
	\lambda^{-\boldsymbol{\mu}\cdot\boldsymbol{p}_i/q_i}
	\boldsymbol{P}\!\left(
		e^{\boldsymbol{y}_x}(\lambda t)^{\boldsymbol{\gamma}_i},
		\lambda t
		\right)^{\boldsymbol{\mu}}
	\right]
	\right|_{\lambda=0}.
	\label{eq:simple-pole-before-yx-translation}
\end{align}
Translating
\(\widetilde{\boldsymbol{y}}_x
=\boldsymbol{y}_x+\boldsymbol{\gamma}_i\ln t\) and renaming
\(\widetilde{\boldsymbol{y}}_x\) as \(\boldsymbol{y}_x\) (note that this would shift around the conical sectors, but the derivation we did would also apply for shifted cones, or we could have done this shift of variables in the very beginning before any sector decomposition), the
complete simple-pole contribution becomes
\begin{align}
	&t^{-s_{p,i,k}(\boldsymbol{s}_x)}
	A_{i,k}(\boldsymbol{s}_x)
	\notag\\
	&\quad=
	\frac{1}{k!}
	\int_{\mathbb{R}^n}d^n\boldsymbol{y}_x\,
	e^{\boldsymbol{s}_x\cdot\boldsymbol{y}_x}
	\left.
	\left(
	\frac{\partial}{\partial\lambda^{1/(-q_i)}}
	\right)^k
	\left[
	\lambda^{-\boldsymbol{\mu}\cdot\boldsymbol{p}_i/q_i}
	\boldsymbol{P}\!\left(
		e^{\boldsymbol{y}_x}\lambda^{\boldsymbol{\gamma}_i},
		\lambda t
		\right)^{\boldsymbol{\mu}}
	\right]
	\right|_{\lambda=0}.
	\label{eq:simple-pole-expansion-by-regions-form}
\end{align}
Finally, return to the original variables
\(x_j=e^{y_j}\), for which
\(d^n\boldsymbol{y}_x=d\boldsymbol{x}/\boldsymbol{x}\) and
\(e^{\boldsymbol{s}_x\cdot\boldsymbol{y}_x}
=\boldsymbol{x}^{\boldsymbol{s}_x}\). Then
\begin{align}
	&t^{-s_{p,i,k}(\boldsymbol{s}_x)}
	A_{i,k}(\boldsymbol{s}_x)
	\notag\\
	&\quad=
	\frac{1}{k!}
	\int_{\mathbb{R}_+^n}
	\frac{d\boldsymbol{x}}{\boldsymbol{x}}\,
	\boldsymbol{x}^{\boldsymbol{s}_x}
	\left.
	\left(
	\frac{\partial}{\partial\lambda^{1/(-q_i)}}
	\right)^k
	\left[
	\lambda^{-\boldsymbol{\mu}\cdot\boldsymbol{p}_i/q_i}
	\boldsymbol{P}\!\left(
		\boldsymbol{x}\lambda^{\boldsymbol{\gamma}_i},
		\lambda t
		\right)^{\boldsymbol{\mu}}
	\right]
	\right|_{\lambda=0}.
	\label{eq:simple-pole-expansion-by-regions-x-form}
\end{align}
Note that \(-q_i\) is the least common multiple of the denominators of the components of \(\boldsymbol{\gamma}_i\) (because \(\boldsymbol{v}_i\) is a primitive integer vector), which was \(m_i\) in Eq.~\eqref{eq:introduction-single-region-full-integral}. Also note that in Eq.~\eqref{eq:introduction-single-region-full-integral} the \(\boldsymbol{n}_i\) are the smallest exponents of \(\lambda^{1/m_i}\), but in Eq.~\eqref{eq:simple-pole-expansion-by-regions-x-form} the \(\boldsymbol{p}_i\) are the largest exponents of \(e^{z_i}=r_i^{-1}\propto \lambda^{-1/(-q_i)}\), from which \(\boldsymbol{p}_i=-\boldsymbol{n}_i\) and \(\boldsymbol{p}_i/q_i=\boldsymbol{n}_i/m_i\). Therefore
Eq.~\eqref{eq:simple-pole-expansion-by-regions-x-form} is the same as the \(k\)-th term of Eq.~\eqref{eq:introduction-single-region-full-integral}.

We have shown that the series expansion we derived from the Mellin transform is equal to the series expansion from the method of expansion by regions, so this is a derivation of the method of expansion by regions.

\end{document}